\documentclass[12pt]{article}
\usepackage{amsmath,amsthm,amsfonts,graphicx,epsfig}
\usepackage[usenames]{color}
\numberwithin{equation}{section}
\newcommand{\ket}[1]{\left| #1 \right\rangle}
\newcommand{\bra}[1]{\left\langle  #1 \right|}
\newcommand{\braket}[2]{\langle #1 | #2 \rangle}

\newcommand{\dbar}{\kern-.1em{\raise.8ex\hbox{ -}}\kern-.6em{d}}
  
\def\half{\mbox{$\frac 1 2$}}
\def\quarter{\mbox{$\frac 1 4$}}
\def\sl{SL(2,\mathbb{C})}
\def\one{\mathbb{I}}
\newtheorem{thm}{Theorem}[section]
\newtheorem{cor}[thm]{Corollary}
\newtheorem{lemma}[thm]{Lemma}
\newtheorem{prop}[thm]{Proposition}

\newtheorem{rem}[thm]{Remark}
\newtheorem{definition}[thm]{Definition}
\def \be{\begin{equation}}
\def \ee{\end{equation}}
\def \bea{\begin{eqnarray}}
\def \eea{\end{eqnarray}}

\usepackage[usenames]{color}

\author{J. E.  Avron and O. Kenneth
\\
Department of Physics\\ Technion, 32000 Haifa, Israel}
\begin{document}
\title{Entanglement and the geometry of two qubits}
\date{\today}%
\maketitle
\begin{abstract}
Two qubits is the simplest system where the notions of separable
and entangled states and entanglement witnesses first appear.
We give a three dimensional geometric description of these notions.
This description however carries no quantitative
information on the measure of entanglement. A four dimensional
description captures also the entanglement measure. We give a neat
formula for the Bell states which leads to a slick proof of the
fundamental teleportation identity. We describe optimal
distillation of two qubits geometrically and present a simple
geometric proof of the Peres-Horodecki separability criterion.
\end{abstract}

%%%%%%%%%%%%%%%%%%%%%%%%%%%%%%%
%%%%%%%%%%%%%%%%%%%%section 1
%%%%%%%%%%%%%%%%%%%%%%%%%%%%%%%%%%%%%%%%%%%%%%%%%%%%%%%%%%%%%%%%%%%%
\section{Introduction and overview}\label{sec:intro}
Geometric descriptions of physical notions are often both useful
and elegant. For example, the geometric description of a single
qubit\footnote{D. Mermin \cite{Mermin} advocates the spelling
``Qbit''.} in terms of the Bloch sphere is a natural way of
introducing the notion of a qubit \cite{Nielsen-Chuang} and at the
same time is also a standard tool in the study of the polarization
of photons \cite{Scully}.

Two qubits are the simplest setting where the notion of
entanglement first appears. Our aim is to describe the world of
two qubits geometrically. Algebraically, the world of two-qubits
is associated with $4\times 4$ Hermitian matrices. This is a
linear space of 16 dimensions. The large dimension makes it hard
to visualize. To have a useful geometric description one needs to
introduce appropriate equivalence relations which preserve the
notions one wishes to describe while substantially reducing the
dimension.

The fundamental notion of equivalence in quantum information
reflects the freedom of all parties to independently choose bases
for their Hilbert spaces. For a pair of qubits shared by Alice and
Bob, this freedom is expressed by a pair of $SU(2)$ operations.
Since ${\rm dim}\, SU(2)=3$, this freedom corresponds to a 6
dimensional family of unitary transformation. This reduces the 15
dimensions that describe the (normalized) states of a general 2
qubits state to 9, which is still too large to be really useful
\footnote{There are, however, certain interesting lower
dimensional families  of states for which the reduction is
powerful enough \cite{Horodecki:information}.}.

To further reduce the dimension one can allow Alice and Bob more
freedom. The standard protocols of quantum information, such as
LOCC ({\em Local operation and classical communication})
\cite{Bennett:LOCC} give Alice and Bob an arsenal of local
operation: Besides the local unitary transformations they are also
allowed to make local measurement and to communicate about what
they did and what they got. They are not allowed to exchange
qubits, however. In LOCC they are also not allowed to discard
qubits but are allowed to do so in SLOCC ({\em Stochastic local
operation and classical communication}), \cite{Bennet:SLOCC}. This
makes SLOCC a filtering process.

LOCC and SLOCC do not naturally lead to equivalence relations but
rather to partial order.  For example, it is a fundamental feature
of entanglement, arguably its defining property, that entanglement
can not be created by local operation \cite{Plenio:thermo}
although it can be locally degraded and destroyed.

We therefore need to introduce a different class of operations
that can serve as an equivalence relation. {\em We shall consider
two states as equivalent if each can be prepared from the other
(filtered) with finite probability by local operations.} Unlike
LOCC or SLOCC, this is a symmetric relation, and hence an
equivalence. It restricts the local operations to those
represented by invertible matrices
\cite{Leinaas-Myrheim-Ovrum,Verstraete:svd}. In particular, Alice
and Bob {\em are not} allowed to make projective measurement or
mix pure qubits because these operations are not reversible, even
probabilistically. (More on this, below).

\begin{figure}[ht]
\hskip 4 cm
\includegraphics[width=5cm]{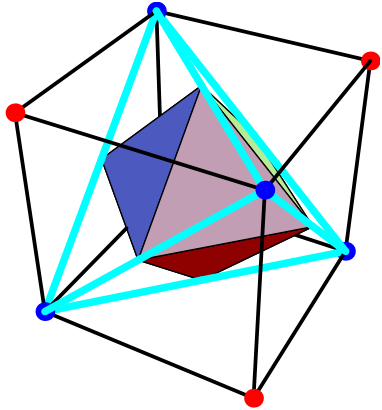}
\caption{The octahedron represents the equivalence class of
separable states. The set of points that lie outside the
octahedron but inside the tetrahedron represent the equivalence
class of entangled states. The set of points that lie outside the
tetrahedron but inside the cube represent entanglement witnesses.
The vertices of the tetrahedron represent the equivalence  class
of pure states. Points related by the tetrahedral symmetry
represent the same equivalence class. }
\label{fig:octa-tetra-cube}
\end{figure}

For describing {\em notions}, such as entanglement and witnesses,
it is convenient to forget about the normalization of states.
This allows one          %%% This will allow us
to describe the world of two qubits in three dimensions
\cite{Leinaas-Myrheim-Ovrum}, as shown in
Fig.~\ref{fig:octa-tetra-cube}. Interestingly, the same figure
appears in various other contexts in quantum information theory.
It first appeared in the Horodeckie`s description
\cite{Horodecki:information} of 2 qubits with maximally mixed
subsystems.  It also appears in the characterization of the
capacity of a single qubit quantum channel
\cite{Ruskai,Ruskai:EBC,Ruskai:king,Ruskai:cp-trace,Zyczkowsky}
and in other contexts
\cite{Avron:visualizing,Werner:comptable_entangl,
Horodecki:PhysRevLett,Tsallis:PhysRevA}.

The 3 dimensional description, beautiful as it is, has weaknesses.
One is that there are certain (fortunately, non-generic) states
that do not seem to fit anywhere in
Fig.~\ref{fig:octa-tetra-cube}. An example is the family of states
where at least one subsystem is pure
    \begin{equation}\label{mixed-pure}
    ({pure})_A\otimes({mixed})_B,\quad ({mixed})_A\otimes({pure})_B,
    \quad ({pure})_A\otimes({pure})_B
    \end{equation}
Another weakness is that the 3 dimensional figure gives no
information on the measure of entanglement:  The distance from the
octahedron does not reflect any of the accepted measures of
entanglement \cite{Plenio-Virmani}. This is a consequence of the
fact that the normalization of states does not matter in the 3
dimensional description.

\begin{figure}[ht]
    \hskip 5 cm
    \includegraphics[width=6cm]{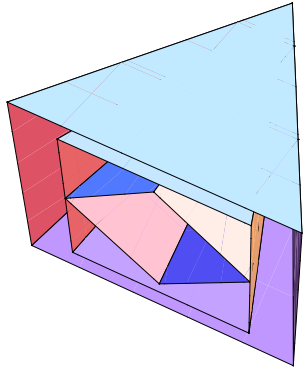}
    \caption{Four dimensional truncated cones describe the LSL equivalence classes of the
    of trace normalized witnesses, entangled and separable states.
    The cross section of the cones, here represented by nested rectangles, are
    the 3 dimensional polyhedra shown in
    Fig.~\ref{fig:octa-tetra-cube}. The largest cone is the cone
    of potential witnesses and the smallest cone is the dual cone
    of separable states. The intermediate cone is the self-dual
    cone of states.  The distance from the cone of separable states is a measure of
    entanglement. The extraordinary families of Eq.~(\ref{mixed-pure})
    are represented by the apex of the cone.}
    \label{fig:cone}
    \end{figure}

To remedy this,  we look at operations where the normalization of
states matters. Specifically, we allow Alice and Bob to act on
their qubits by matrices $M_{A,B}\in \sl$, the group of $2\times
2$ matrices, with complex entries and unit determinant. The
interpretation of this family in terms of measurments shall be
discussed in section \ref{sec:local-operations}. We shall call
this class of operations LSL (for local, special and linear). LSL
allows for a geometric description of the measure of entanglement.
The price one pays is that  one needs to go to 4 dimensions. The
geometric picture that emerges is illustrated in
Fig.~\ref{fig:cone}, showing three nested cones. The largest cone
is the cone of potential witnesses, whose cross section is the
cube in Fig.~\ref{fig:octa-tetra-cube}. The boundary of the cone
is special in that it is cohabited by two inequivalent families:
The ordinary and the extraordinary. This makes it pathological
\footnote{When distinct points can not be separated a space is
non-Hausdorff.}. It may seem odd that the world of 2 qubits, which
is a simple linear space in 16 dimensions, becomes pathological
when viewed in terms of its equivalence classes. A useful analogy
is the partitioning of (the connected) Minkowsky space-time to the
(disconnected) equivalence classes of time-like, light-like and
space-like vectors.

The four dimensional description {\em is} faithful to the measure
of entanglement.  More precisely, the concurrence,
\cite{Wootters:concurrence,Wootters:review}, is the distance from
the cone of separable states, the smallest nested cone. In
particular, states represented by points near the apex of the cone
have very little entanglement.
%%%%%%%%%%%%%%%%%%%%%%%%%%%%%%%%%%%%%%%%%%%%%%%%%%%%%%%%%%%%%%

Many things will have to be left out. Among them: the notions of
``entanglement of formation'', ``entanglement cost''
\cite{Bennett:LOCC,Bennet:SLOCC,Hayden};  ``bound entanglement''
\cite{Horodecki:bound-ent}, ``entanglement persistence''
\cite{Briegel-Raussendorf}, multiparities entanglement, GHZ states
\cite{GHZ} and the different entanglement measures
\cite{Plenio-Virmani}. Comprehensive reviews of entanglement, with
extensive bibliography, are
\cite{zyczkowski-2006,zyczkowski-bengtsson,Horodecki:review}.

%%%%%%%%%%%%%%%%%%%%%%%%%%%%%%%%%%%%%%%%%%%%%%%%
%%%%%%%%%%%%%%%%%%%%%%%%%%%%%%%%%%%%%%%%%%%%%%%

%%%%%%%%%%%%%%%%%%%%%%%%%%%%%%%
%%%%%%%%%%%%%%%%%%%%%%%%%%%%%%%
%%%%%%%%%%%%%%%%%%%%% section 2
%%%%%%%%%%%%%%%%%%%%%%%%%%%%%%%
%%%%%%%%%%%%%%%%%%%%%%%%%%%%%%%

%%%%%%%%%%%%%%%%%%%%%%%%%%%%%%%
\section{Bell states}

The mothers of all entangled states are the four Bell states
\cite{Braunstein}, commonly denoted by $\ket{\beta_\mu}$, here
chosen to be
    \bea
   {\sqrt{2}} \ket{\beta_0}={\ket{00}+\ket{11}},&\quad&
     {\sqrt{2}} \ket{\beta_1}={\ket{00}-\ket{11}}\nonumber\\
     {\sqrt{2}}\ket{\beta_2}=\ket{01}+\ket{10},&\quad&
    i\sqrt 2 \ket{\beta_3}=\ket{01}-\ket{10}
    \eea
The (isotropic) singlet is then $\ket{\beta_3}$. It is not a
coincidence that the number of Bell states coincides with the
number of Pauli matrices $\sigma_\mu$, (with $\sigma_0$ the
identity).
    \begin{prop}
    The Pauli matrices give the unitary map from the computational basis,
    $\ket{a}\otimes \ket{b}$, with $a,b$
    binary, to the Bell basis $\ket{\beta_\mu},\ \mu\in \{0,1,2,3\}$. Explicitly:
        \be\label{eq:bell-states}
     \sqrt{2}\ket{\beta_\mu}=(\sigma_\mu)_{ab}\ket{a}\otimes\ket{b},\quad
     \sqrt{2}\ket{a}\otimes\ket{b}= (\sigma_\mu^t)_{ab}\ket{\beta_\mu}
    \ee
    Summation over repeated indices is implied and the
    4 Pauli matrices are chosen as
    \bea\label{eq:pauli-mat}
    \sigma_0=\left(%
        \begin{array}{cc}
         1 & 0 \\
          0 & 1 \\
        \end{array}%
    \right), \quad
    \sigma_1=\left(%
        \begin{array}{rr}
          1 &  0 \\
          0 & -1 \\
        \end{array}%
    \right),\nonumber \\
    \sigma_2=\left(%
        \begin{array}{cc}
         0 & 1 \\
          1 & 0 \\
        \end{array}%
    \right),\quad
    \sigma_3=\left(%
        \begin{array}{rr}
         0 & -i \\
          i & 0 \\
        \end{array}%
    \right).
    \eea
     \end{prop}
Note that the anti-symmetric Pauli matrix is $\sigma_3$, (rather
than the more common choice $\sigma_2$), a choice also made in
\cite{Horodecki:review}.
%%%%%%%%%%%%%%%%%%%%%%%%%%
    \begin{prop}\label{prop:permutation}
     The basic two-qubit operations,
     $\sigma_\alpha\otimes\sigma_\beta$,
    act on the Bell states as a permutation, (up to phase
    factors) and generate the the symmetry group of the tetrahedron.
    \end{prop}
Proof: From Eq.~(\ref{eq:bell-states})
    \be\label{eq:local-bell}
    \sigma_\alpha\otimes\sigma_\beta\ket{\beta_\mu}=\frac 1{\sqrt{2}}
    (\sigma_\alpha\sigma_\mu\sigma_\beta^t)_{ab}\ket{ab}
    \ee
Since
    \be\label{eq:sigma-algebra}
    \sigma_\mu\sigma_\nu=\left\{%
\begin{array}{ll}
    i\epsilon_{\mu \nu k} \sigma_k, & \hbox{$\mu,\nu,k \in \{1,2,3\}$;} \\
    \sigma_\mu, & \hbox{$\nu=0$;} \\
    \sigma_\nu, & \hbox{$\mu=0$.} \\
\end{array}%
\right.
    \ee
We see that $\sigma_\alpha\otimes\sigma_\beta$ just permutes the
Bell states (up to phase factors). Since $\sigma_j\otimes\sigma_0$
interchange $\ket{\beta_0}\leftrightarrow \ket{\beta_j}$, they
generate the permutation group of the four Bell states, $S_4$. It
is a fact that the tetrahedral group coincides with $S_4$.
 \hfill$\square$

The Bell projections play a key role in what we do.
%They turn out to have the following {\em canonical form}
    \begin{prop}
    The Bell projection  $P_\mu= \ket{\beta_\mu}\bra{\beta_\mu}$
    (no summation over $\mu$, of course) have the  form
    \begin{eqnarray}\label{eq:bell}
    4P_0&=&\sigma^{\otimes 2}_0+\sigma^{\otimes 2}_1
    +\sigma^{\otimes 2}_2-\sigma^{\otimes 2}_3\ ,\nonumber\\
     4P_1&=&\sigma^{\otimes 2}_0+
    \sigma^{\otimes 2}_1-\sigma^{\otimes 2}_2+\sigma^{\otimes 2}_3\ ,
    \nonumber\\
      4P_2&=&\sigma^{\otimes 2}_0-\sigma^{\otimes 2}_1
    +    \sigma^{\otimes 2}_2+\sigma^{\otimes 2}_3\ ,        \\
     4P_3&=&
    \sigma^{\otimes 2}_0-\sigma^{\otimes 2}_1-\sigma^{\otimes 2}_2-\sigma^{\otimes 2}_3\nonumber
    \end{eqnarray}
where we denote
    \be
    \sigma_\mu^{\otimes 2}=\sigma_\mu\otimes\sigma_\mu
    \ee
    \end{prop}
Proof: From Eq.~(\ref{eq:bell-states}) one finds (no summation
over $\mu$ below),
    \bea\label{eq:bell-projection}
   8P_\mu= &=& 4
    (\sigma_\mu)_{ab}(\sigma_\mu)_{dc}\ket{a}
    \bra{c}\otimes\ket{b}\bra{d}\nonumber \\
   &=&
    (\sigma_\mu)_{ab}(\sigma_\mu)_{dc}(\sigma_\beta)_{ca}
    (\sigma_\alpha)_{db}\,\sigma_\beta\otimes\sigma_\alpha
     \\
     &=&
    Tr\,\big(\sigma_\beta\sigma_\mu
    \sigma_\alpha^
    t\sigma_\mu\big)\,\sigma_\beta\otimes\sigma_\alpha\nonumber
    \eea
In the second line we used
    \be
    2\ket{a}\bra{c}=Tr\big(
    \ket{a}\bra{c}\sigma_\beta\big)\ \sigma_\beta=(\sigma_\beta)_{ca}\sigma_\beta.
    \ee
Since the Pauli matrices either commute or anti-commute,
are either symmetric or antisymmetric, and are mutually
orthogonal we have (no summation over $\mu$ here)
    \be
     Tr\,\big(\sigma_\beta\sigma_\mu
    \sigma_\alpha^t\sigma_\mu\big)=\pm
    Tr\,\big(\sigma_\beta
    \sigma_\alpha^t\big)=\pm
    Tr\,\big(\sigma_\beta
    \sigma_\alpha\big)=\pm 2\delta_{\alpha,\beta}
    \ee
Hence, only the diagonals survive in
Eq.~(\ref{eq:bell-projection}).\hfill$\square$

\subsection{Teleportation}

Bell states can be used to teleport \cite{Teleportation} an
unknown qubit. This is a consequence of the following
teleportation lemma\footnote{The formula seems to be related to a
formula in \cite{zcachor}.} :
    \begin{lemma}  Let $\ket{\psi}$ be a single qubit pure state.
    Then the following identity holds
    \be
    2\ket{\psi}\otimes \ket{\beta_\mu}= \ket{\beta_\nu}\otimes
    \ket{\sigma_\mu^t\sigma_\nu\psi}
    \ee
    \end{lemma}
    Proof: From Eq.~(\ref{eq:bell-states}):
    \bea
     2 \ket{\psi}\otimes \ket{\beta_\mu}&=&\sqrt 2\psi_c (\sigma_\mu)_{ab}\ket{c}\otimes\ket{a}\otimes\ket{b}
    \nonumber \\
    &=&\psi_c (\sigma_\mu)_{ab}(\sigma_\nu^t)_{ca}\ket{\beta_\nu}\otimes\ket{b}
    \nonumber \\
    &=&(\sigma_\mu^t\sigma_\nu\psi)_b \ket{\beta_\nu}\otimes\ket{b}
     \\
    &=&\ket{\beta_\nu}\otimes
    \ket{\sigma_\mu^t\sigma_\nu\psi}\nonumber\quad\quad\quad\quad\square
    \eea

The identity has the following physical interpretation: The left
hand side describes the situation where Alice has the (unknown)
qubit $\ket{\psi}$ and shares with Bob the Bell state
$\ket{\beta_\mu}$. The right hand side describes the superposition
of the following situations: Alice pair of qubits is in one of the
four Bell states while Bob's qubit is a unitary transformation of
$\ket{\psi}$. Alice can then measure in the Bell basis and tells
Bob which Bell state she finds.  Bob then performs the unitary
operation $\sigma_\nu\sigma_\mu^t$ on his  qubit to retrieve
$\ket{\psi}$.

\subsection{The CHSH Bell inequalities}

The Bell states have the distinguished property that they give
maximal violation of the CHSH Bell inequalities \cite{CHSH}. Bell
inequalities \cite{bell} show that quantum mechanics can not be
simulated by classical probability theory
\cite{Bell-speakable,Mermin-Boojums,Peres:book,Tsirelson,Holevo}.
This bit of theory follows simply from the formulas above for the
Bell projections, as we now outline.

Let us denote by $a_{1,2}$ the result of Alice measurement of
$\sigma_{1,2}$ and by $b_{+,-}$ the result of Bob measurement of
$(\sigma_1\pm \sigma_2)/\sqrt 2$. All these measurements are
dichotomic and yield only $\pm 1$. Any assignment of $\pm 1$ to
the corresponding 4 measurements yields
    \be\label{eq:chsh}
    -2\le a_1(b_++b_-)+ a_2(b_+-b_-)\le 2
    \ee
The same inequality must also hold on the average for any ensemble of classical systems.
This is the CHSH Bell
inequality\cite{CHSH,Nielsen-Chuang,Peres:book}.

Quantum mechanics is inconsistent with this inequality. To see
this define the Bell operator \cite{Braunstein} to be the
observable corresponding to Eq.~\ref{eq:chsh}:
    \bea\label{eq:bell-operator}
    \mathbf{B}&=& \sigma_1\otimes\left(\frac{\sigma_1+\sigma_2}
    {\sqrt 2}+\frac{\sigma_1-\sigma_2}{\sqrt 2}\right)+
    \sigma_2\otimes\left(\frac{\sigma_1+\sigma_2}
    {\sqrt 2}-\frac{\sigma_1-\sigma_2}{\sqrt 2}\right)\nonumber \\
    &=& \sqrt 2\big(\sigma^{\otimes 2}_1+\sigma^{\otimes
    2}_2\big)=
    2\sqrt 2\big( P_0-P_3\big)
    \eea
Clearly, $\ket{\beta_0},\ \ket{\beta_3}$ are eigenstates of
$\mathbf{B}$ with eigenvalues $\pm 2\sqrt 2$ and hence violate
 Eq.~(\ref{eq:chsh}). The
probabilistic aspect of quantum mechanics can not be attributed to
a classical probabilistic source that prepares the qubits of Alice
and Bob.

%%%%%%%%%%%%%%%%%%%%%%%%%%%%%%%
%%%%%%%%%%%%%%%%%section 3%%%%%

\section{Separable states}

In classical probability theory, random variables  $x$ and $y$
are independent when their joint probability distribution is
a product $P_A(x) P_B(y)$. Any joint probability distribution
$P_{AB}(x,y)$ can be trivially written as a convex combination of product
distributions:
    \be
    P_{AB}(x,y)=\sum_{\alpha,\beta} P_{AB}(\alpha,\beta)\,
    \delta_{x,\alpha}\delta_{y,\beta}
    \ee
where $P_{AB}(\alpha,\beta)$ are thought of as weights and the two
delta functions as probability measures.

This is not true in quantum mechanics \cite{Holevo}. A state
$\rho$, a positive matrix with unit trace, is the analog of a
probability measure. A state of the form $\rho_A\otimes \rho_B$
describes the situation where Alice`s and Bob`s qubits are
uncorrelated. However, it is {\em not true} that all states can be
written as convex combinations of uncorrelated states.  The states
that can be written in this way are called {\em separable}
\cite{werner,Horodecki:review}.
    \begin{definition}
    A (normalized) state $\rho_s$ is separable if
    \begin{equation}\label{eq:separable}
    \rho_s=\sum_{n=1}^N \,p_n \, { \rho}_A^{(n)}\otimes {\rho}_B^{(n)},
    \end{equation}
    with $p_n\ge 0$ probabilities and $\rho^{(n)}_{A,B}$ positive operators
    with normalized trace.
    A state $\rho\ge 0$ which is not separable is {\em entangled}.
    \end{definition}
Clearly, the unnormalized separable states make a convex cone
contained in the cone of all positive (unnormalized) states.

Separable states can be interpreted as mixtures of uncorrelated
states where Alice and Bob rely on a common probability
distribution, $p_n$ to create the mixture. This correlates Alice
and Bob. Such correlation never violate Bell inequalities.  For
the CHSH this can be seen from the fact that for any product state
    \be
     \left|Tr(\mathbf{B} \rho_A\otimes\rho_B)\right|=\left|a_1 (b_++b_-) +
    a_2(b_+-b_-)\right|\le |b_++b_-| +     |b_+-b_-|\le 2\nonumber
    \ee
where now
    \be
    |a_{1,2}|=|Tr(\sigma_{1,2} \rho_A)|\le 1, \quad
    |b_{+,-}|=\left|Tr\left(\frac{\sigma_1\pm\sigma_2}{\sqrt 2}
    \rho_B\right)\right|\le 1
    \ee
This result  extends to separable states by convexity.

States that violate a Bell inequality are necessarily entangled.
However, there are lots of entangled states that {\em do not}
violate any CHSH inequality. (The equivalence classes of states
that satisfy the CHSH inequality and their visualization is given
in \cite{Avron:visualizing}.)

There are no known general conclusive tests of separability.
However, for 2 qubits the Peres-Horodecki partial transposition test
\cite{Peres:test,Horodecki:iff} gives a simple spectral test of
separability. To describe this test we first explain the notion of
partial transposition for 2 qubits.

Any observable (Hermitian matrix) in the space of 2 qubits can be
written as
    \begin{equation}\label{eq:A-lorentz-comp}
    A={\sf A}^{\mu \nu}\sigma_\mu\otimes\sigma_\nu, \quad  {\sf A}^{\mu
    \nu}\in \mathbb{R}
    \end{equation}
For reasons that shall become clear later we call ${\sf A}^{\mu
\nu}$ the (contravariant) {\em Lorentz components} of $A$. The
partial transpose of $A$, which we denote by $A^P$, is
    \be\label{eq:partial-transpose}
    A^P={\sf A}^{\mu \nu}\sigma_\mu\otimes\sigma_\nu^t=
    ({\sf A}^P)^{\mu \nu}\sigma_\mu\otimes\sigma_\nu
    \ee
Observing that only $\sigma_3$ is anti-symmetric we see that the
partial transpose, when expressed in terms of the Lorentz
components, takes the form
     \be\label{eq:partial-transpose-matrix}
    ({\sf A}^P)^{\mu \nu}=\left\{%
\begin{array}{ll}
    {\sf A}^{\mu \nu}, & \hbox{$\nu \neq 3$;} \\
     -{\sf A}^{\mu \nu}, & \hbox{$\nu=3$.} \\
\end{array}%
\right.
    \ee
The Peres-Horodecki test is \cite{Peres:test,Horodecki:iff}
    \begin{thm}\label{thm:ph}
    A 2 qubit state $\rho\ge 0$ is separable iff $\rho^P\ge 0$
    \end{thm}
Proof: The ``if'' part is easy: If $\rho$ is separable, it can be
written as in Eq.~(\ref{eq:separable}). Since $\rho_B\ge 0$ implies
that also $\rho^t_B\ge 0$, one has that $\rho^P$, being a convex
combination of positive operators, is also positive. The ``only
if'' part requires more preparations. We shall give a simple
geometric proof in section \ref{sec:peres}. \hfill $\square$
%%%%%%%%%%%%%%%%%%%%%%%%%%%%%%%
%%%%%%%%%%%%%%%%%%%%%%%%%%%%%%%

\section{Witnesses}
\begin{figure}[ht]
    \hskip 3.5 cm
    \includegraphics[width=6cm]{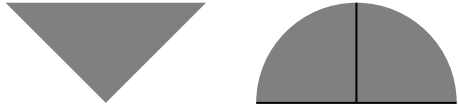}
    \caption{A cone with opening angle $\theta$ is dual
    to the cone with opening angle $\pi-\theta$. The quadrant,
    (left) is self-dual. The half-plane (right) is dual to
    the positive half-line. Here the usual scalar product of vectors corresponds
    to the trace in Eq.~(\ref{eq:witness}). }
    \label{fig:duality}
    \end{figure}

Witnesses are observables which can give evidence that a state is entangled.
For our present purposes it is convenient to slightly widen this notion and to
allow for witnesses which are in a sense trivial.
We therefore define the cone of {\em potential witnesses} as follows:
    \begin{definition}\label{def:evidence}
    The dual cone\footnote{The notion of dual cones is illustrated in Fig.~\ref{fig:duality}.} to the cone of separable states shall be called
    the cone of {\em potential witnesses}. Explicitly, it is
    the collection of all observables $W$ such that
        \be\label{eq:witness}
        Tr({ W}\rho_s)\ge 0,
        \ee
    for all separable states $\rho_s$. We shall call $-Tr(W\rho)$
    {\em the (entanglement) evidence}.
    \end{definition}

A potential witness is called simply a witness iff there exist
some (necessarily entangled) state $\rho$ such that $Tr({
W}\rho_s)< 0$. The witness then give conclusive evidence that
$\rho$ is entangled. In fact by standard duality arguments
\cite{Rockafellar} the definition implies
    \begin{prop}\label{prop:witness}
     Any entangled state, $\rho$, has a witness $W$  giving
    positive evidence, i.e.
    \be
    -Tr(W \rho)>0
    \ee
    \end{prop}

The set of potential witnesses (unlike witnesses proper) is a
convex set. A potential witness $W$ may not give positive evidence
for any state $\rho$. Clearly, this will be the case whenever $W\ge 0$.  %% is a positive operator %%
Thus the cone of potential witnesses contains the cone of states:
    \be\label{ccc}
    Potential \ Witnesses\supset States \supset Separables
    \ee

Observe that since $\rho = \one$ is clearly a separable
(un-normalized) state it follows that any potential witness $ W$
has  $  Tr \, W \geq 0$. Moreover, the following holds:
    \begin{thm}
    For any potential witness $W$, not identically zero,
    \be\label{eq:trace-witness}
    Tr \,  W > 0  .
    \ee
    In particular, witnesses, like states,  may be normalized to have unit trace.
    \end{thm}
Proof: Note that the elements
$\bra{\varphi\otimes\psi}W\ket{\varphi\otimes\psi}$ suffice to
determine all other matrix elements of $W$. This may be verified
by considering the case
${\varphi\otimes\psi}=(\varphi_1+e^{i\alpha}\varphi_2)\otimes(\psi_1+e^{i\beta}\psi_2)$
for all $\alpha$'s and $\beta$'s. For any $ W\neq0$ one can
therefore always find a normalized product state
$\ket{\varphi_0}\otimes\ket{\psi_0}$ such that
$\bra{\varphi_0\otimes\psi_0}W\ket{\varphi_0\otimes\psi_0}> 0$.
Complete this to an orthonormal product base
$\{\ket{\varphi_a}\otimes\ket{\psi_b}\}_{a,b=0,1}$. Since
    \be
    Tr \, W
    =\bra{\varphi_a\otimes\psi_b} W\ket{\varphi_a\otimes\psi_b}\, ,
    \ee
(summation implied), has no negative terms one concludes the
strict inequality. \hfill $\square$

It remains to demonstrate that there indeed are entangled states
or, equivalently, that the inequality Eq.~(\ref{ccc}) is strict.
An example for a witness that is not a positive operator is the
the swap, which exchanges the qubits of Alice and Bob:
    \be
    S\ket{\psi}\otimes \ket{\phi}=\ket{\phi}\otimes \ket{\psi}
    \ee
    \begin{prop}
    The swap has the following properties
    \begin{enumerate}
    \item $S$ is positive on separable states
    \item $S=P_0+P_1+P_2-P_3$ gives positive evidence that the singlet
    is entangled
    \item The swap is the partial transpose of  a Bell projection: $ S= 2 P_0^P$
    \end{enumerate}
    \end{prop}
    Proof: That $S$ is positive on all pure product states follows
    from:
    \begin{equation}\label{exchange}
    {}_A\bra{\psi}\otimes {}_B\bra{\phi}S\ket{\psi}_A\otimes
    \ket{\phi}_B=\braket{\psi}{\phi}_A\braket{\phi}{\psi}_B=|\braket{\psi}{\phi}|^2
    \end{equation}
It is then positive on all separable states by convexity and so
belongs to the cone of potential witnesses. This proves 1. Part 2
follows by noting that the Bell states are eigenvectors of the
swap. Part 3 follows from the observation that swap can be written as
$S=\ket{ab}\bra{ba}$, while $P_0=\half \ket{aa}\bra{bb}$.
\hfill$\square$

It is, of course, not a coincidence that the swap is a witness of
a Bell state. In Fig. \ref{fig:octa-tetra-cube} Bell states are
represented by the (blue) dots at the vertices of the tetrahedron
and the witnesses by the red dots at the corners of the cube
obtained by reflection about the 3 axis. We shall see in Corollary
\ref{thm:optimal-witness} below, that the swap is, in fact, an
optimal witness.

\section{Equivalence and Local operations}\label{sec:local-operations}

We shall consider equivalence classes where $\rho$ and $\rho^M$
are considered equivalent provided
    \be\label{eq:LSL}
    \rho\mapsto \rho^M= M\rho M^\dagger, \quad M=M_A
    \otimes M_B
    \ee
with $M_{A,B}$ taking values in the groups
    \be\label{eq:groups}
    SU(2)\subset\sl\subset GL(2,\mathbb{C})
    \ee
The equivalence clearly preserve the positivity and the
separability of states but, in general, not its normalization.

$M_{A,B}\in \sl$ will turn out to be our main tool and shall be
designated by the  acronym LSL for {\em local, special (-unit
determinant) and linear}.

The linear maps in Eq.~(\ref{eq:LSL}) with $M_{A,B}\in \sl $ or
$GL(2,\mathbb{C})$ do not represent, in general, operations that
Alice and Bob can perform on their qubits. Legitimate {\em quantum
operation} are positivity preserving and trace non-increasing
\cite{Peres:book,Nielsen-Chuang}. This means that $M$ in
Eq.~(\ref{eq:LSL}) must satisfy $M^\dagger M\le 1$. Quantum
operation with $M^\dagger M< 1$ are interpreted as a generalized
measurement, aka POVM
\cite{Davies,Holevo,Peres:book,Nielsen-Chuang,brandt:434}.

When $ M_{A,B}\in\sl$ or $ GL(2,\mathbb{C})$, $M^\dagger M\not\le
1$ so the linear map in Eq.~(\ref{eq:LSL}) do not represent
legitimate quantum operations. Nevertheless with every such group
element $M$ we can associate the bona-fide POVM element
    \be
    M\mapsto \frac M{\|M\|}
    \ee
The corresponding measurement filters
\cite{Gisin-filters,Gisin:filter} the state
     \be\label{eq:filtered-state}
    \rho \mapsto  \frac{M\rho M^\dagger}{Tr (M\rho M^\dagger) }
    \ee
Filtering wastes a fraction of the qubits which Alice and Bob need
to discard.  Indeed, the filtration succeeds with probability
    \be\label{eq:prob-success}
    p(\rho)= \frac{Tr (M\rho M^\dagger)}{\|M\|^2}\le 1
    \ee
(With $M_{A,B}\in SU(2)$ the ``filtration'' succeeds with
probability one, but with $M_{A,B}\in \sl,  GL(2,\mathbb{C})$
not.) Alice and Bob need to communicate over a classical channel,
so they both keep only the qubits that pass the tests. This makes
LSL a special case of SLOCC.

The equivalence classes introduced in Eq.~(\ref{eq:LSL}) therefore
admit the interpretation that two states are equivalent provided
each can be filtered from the other and the filtration succeeds
with finite  probability.  This imposes a restriction on what
Alice and Bob are allowed to do. In particular, mixing   %%and purifying
is not an admissible operation. This is easily seen from
the fact that $M$ preserve the rank of $\rho$ and so maps pure
states to (unnormalized) pure states.

\subsection{The equivalence classes of a single qubit}

To appreciate the various notions of equivalence introduced above
consider a single qubit. Any single qubit state $\rho$, can be
identified with a (real) 4-vector ${\sf r}_\mu$
    \be\label{eq:lorentz-1-qubit}
    \rho={\sf r}^\mu\sigma_\mu,
    \ee
We shall refer to ${\sf r}^\mu$ as the (contravariant) ``Lorentz
components'' of $\rho$. States admit the following simple
geometric characterization:
    \begin{lemma}\label{lem:light-cone}
    The 4-vector ${\sf r}^\mu$ represents an (un-normalized)
    state iff it lies in the forward light-cone.
    Pure states are light-like. Normalized
    states lie on the time slice $t=\half$.
    \end{lemma}
Proof: This easily follows from
    \begin{equation}\label{det-trace}
    \det { \rho }= {\sf r}^\mu {\sf r}^\nu\eta_{\mu\nu}, \quad Tr { \rho}=2{\sf r}^0
    \end{equation}
where $\eta$ is the Minkowsky metric tensor,
$\eta=diag(1,-1,-1,-1)$. Positivity $\rho> 0$ requires that both the trace
and determinant are non-negative. The 4-vector ${\sf r}^\mu$ must
then lie in the forward light cone. A pure state, being rank one,
has $\det \rho=0$ and is represented by a light-like vector. A
normalized state has $Tr\rho=2{\sf r}^0=1$ and so lies on the
fixed time-slice.\hfill $\square$

It follows from Eq.~(\ref{det-trace}) that if ${ \rho}^M=M{
\rho}M^\dagger$, with $M\in \sl$ then the Lorentz component indeed
transform like a vector under Lorentz transformation
\begin{equation}\label{lorentz}
\left({\sf r}^M\right)^\mu={(\Lambda_M)^{\mu}}_\nu {\sf r}^\nu
\end{equation}
where $\Lambda_M\in SO_+(1,3)$ is an (orthochronos) Lorentz transformation.
%% If $M\in SU(2)$ then the corresponding spatial vector just rotates.
If $M\in SU(2)$ then it just rotates the spatial part of the 4-vector.

   \begin{figure}[ht]
   \vskip 1 cm
    \hskip 4 cm
    \includegraphics[width=5cm]{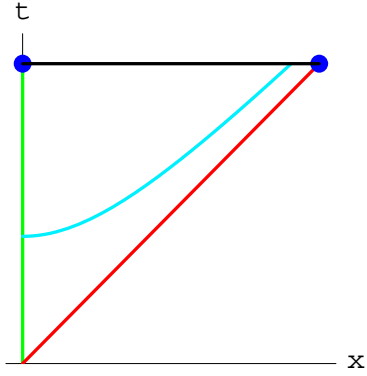}
    \caption{The light cone is represented by the diagonal (red) line.
    The equivalence classes of the $SU(2)$ (normalized) states are represented by the
    horizontal (black) line at fixed time. The $GL(2,\mathbb{C})$ equivalence
    classes are the two (blue) dots. The dot on the time axis represents
    the equivalence class of mixed states and the other, on the light cone,
    the pure states. The $\sl $ equivalence classes are represented by the vertical green line.
    Light-like vectors can be brought to this line only through infinite boost.
    Points representing the $SU(2)$ and $\sl $ equivalence classes that lie on the same interval,
    $t^2-x^2=const$, (cyan) have identical entropies.} \label{fig:qubit}
    \end{figure}

It is instructive to compare the equivalence classes associated
with normalized states of a single qubit with $M$ taking values in
the three groups $ SU(2), \sl, GL(2,\mathbb{C})$, shown in
Fig.~\ref{fig:qubit}.
\begin{itemize}
\item $SU(2)$ acts as spatial rotations and can be used to map any
normalized state to the $x-t$ plane at time slice $t=\half$.

\item $\sl $ acts as a Lorentz transformation and can be used to
transform any time-like vector to the time-axis and any light-like
vectors to any other light-like vector. This means that the LSL
equivalence classes are represented by the semi-open interval
$(0,\half]$ and a point. It is natural to close the interval by
gluing the point to the origin, (since a light-like vector can be
transformed to the origin of the time axis in the limit of
infinite boosts). The  $\sl $ and $SU(2)$ equivalence classes of
{\em normalized} one qubit states are then in 1-1 correspondence.

\item The $GL(2,\mathbb{C})$ equivalence classes, however, are
represented by two points: One representing all pure states
(light-like vectors) and one representing all mixed states.
\end{itemize}

For a normalized qubit the von Neumann entropy, $H(\rho )=-Tr
(\rho \log_2 \rho)$, is uniquely determined by $\det \rho$. To see
this express $\lambda$ be the large eigenvalue of $\rho$, as
$2\lambda= 1 +\sqrt{1-4 \det \rho}$. Then
    \be\label{eq:entropy}
   H(\rho ) =-h(\lambda )-h(1-\lambda),\quad h(\lambda )= \lambda \log_2 \lambda \,.
    \ee
Since $\det \rho$ is preserved by $\sl$ we see that LSL preserves
the information on the entropy of the state (provided it is not
renormalized). $GL(2,\mathbb{C})$ on the other hand, does not
distinguish between mixed states with different entropies.

\subsection{Equivalence classes of two qubit pure states}\label{sec:pure-states}

Two-qubits pure states are conveniently represented either in the
computational basis or in the Bell basis:
    \be\label{eq:psi-A}
    \ket{\psi}=Y_{ab}\ket{a}\otimes\ket{b}=\xi^\mu\ket{\beta_\mu},\quad
     \sqrt 2\, Y=\xi^\mu\sigma_\mu
    \ee
Summation over repeated indices is implied; $a,b\in \{0,1\}$ and
$\mu\in\{0,1,2,3\}$. The state is normalized if $Tr \,(YY^\dagger)
=\xi_\mu \xi_\mu^*=1$. This means that the pure states are
described by the seven-sphere $S^7$
\cite{zyczkowski-2006,Mosseri}.

Local transformations take $\ket{\psi}$ to
      \be\label{eq:lsl-pure-state}
      M_A\otimes M_B\ket{\psi}=
      (M_AYM_B^t)_{ab}\ket{a}\otimes\ket{b}
      \ee
We see that $\det Y$ is invariant under the action of
$M_{A,B}\in\sl$. As we shall see in the next section, $|\det Y|$
is a measure of the entanglement. This makes the entanglement an
LSL invariant.

\subsection{ Entanglement distillation of pure states}

The entanglement of a pure bi-partite normalized state is {\em defined} as
the von Neumann entropy of either of its subsystems:
    \be\label{ee}
    e(\ket{\psi})= H(\rho_A)= H(\rho_B)
    \ee
In the case at hand, where $\ket{\psi}$ is given by the matrix $Y$
of Eq.~(\ref{eq:psi-A}),
    \be\label{yy}
    \rho_A=YY^\dagger,\quad
    \rho_B=Y^\dagger Y.
    \ee
It follows from Eq.~(\ref{yy}) that
    \be
    \det\rho_A=\det\rho_B= |\det Y|^2
    \ee
By Eq.~(\ref{eq:entropy}) $\det \rho_A$ determines the entropy of
Alice's qubit. It follows that the measure of entanglement is
uniquely determined by $\det Y$. Moreover, since $\det Y$ is
invariant under LSL by Eq.~(\ref{eq:lsl-pure-state}), we see that
LSL is a useful equivalence not just for describing the {\em
notion} of entanglement, but also for describing its {\em
measure}.

\begin{figure}[ht]
\hskip 4 cm
\includegraphics[width=6cm]{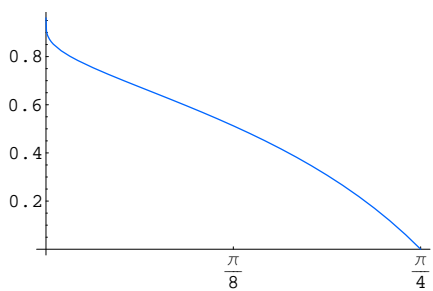}
\caption{The relative dissipation of the entanglement, $(e-p)/e$,
for filtering Bell states from the states in Eq.~(\ref{eq:schmidt})
as a function of $\theta$. The entanglement is $e$ and the
probability of filtering a Bell state is $p$. The graph expresses
the inequality in Eq.~(\ref{eq:dissipation}). }
\label{fig:entanglement}
\end{figure}

One must distinguish between the (mathematical) fact that the
information on the measure of entanglement is preserved under LSL
and the (physical) principle that local operations
%can only
dissipate entanglement \cite{Plenio:thermo}. The difference comes
from the way both treat the issue of normalization. An an example,
consider
%the, in general partially entangled state
    \be\label{eq:schmidt}
    \ket{\psi_\theta}=\cos \theta \ket{00}+\sin\theta\ket{11}, \quad
    0\le\theta\le \pi/4
    \ee
The LSL operation
    \be
    M=\big(\sqrt{\tan\theta} \ket{0}\bra{0}+\sqrt{\cot\theta}
    \ket{1}\bra{1}\big)\otimes \one
    \ee
filters from it the fully entangled unnormalized Bell state
$\sqrt{\sin\theta\cos\theta}\ket{\beta_0}$. The information on the
original measure of entanglement sits in the normalization. At the
same time the corresponding quantum operation dissipates
entanglement. By Eq.~(\ref{eq:prob-success}), the operation
succeeds with probability
    \be
    p(\ket{\psi_\theta})=2\sin^2\theta
    \ee
Using the fact $e(\ket{\beta_0})=1$ one ends up with less
entanglement:
    \be\label{eq:dissipation}
    p(\ket{\psi_\theta})= p(\ket{\psi_\theta}) e(\ket{\beta_0}) \le e(\ket{\psi_\theta})
    \ee
in accordance with the principle that local operations dissipate
entanglement, see Fig. \ref{fig:entanglement}.

\section{Duality of states and observables}

\subsection{Contragradient actions}\label{sec:time-reversal}
States $\rho$ and observables $ W$ naturally live in dual spaces
since pairing the two, $Tr\,(\rho W)$, gives a number. It is both
natural and convenient to define the operations so that they act
on states and witnesses in a way that respects their duality.
Namely:
    \bea\label{eq:observables}
    \rho\mapsto \rho^M
    &=&M^\dagger \rho M\nonumber \\
     W\mapsto  W^M
    &=& M^{-1} W (M^{-1})^\dagger
    \eea
If $M$ is unitary, $M^{-1}=M^\dagger$, then states and observables
transform the same way, but when $M$ is only invertible, they do
not. With this choice $\rho  W$ undergoes a similarity
transformation
    \begin{equation}\label{eq:AW^diamond}
    \rho  W\mapsto \rho^M( W)^M=M^\dagger\,(\rho  W)\, (M^\dagger)^{-1}
    \end{equation}
and $Tr(\rho W)$ is invariant.

When the local operations are taken from $\sl$, there is a map,
the tilde map, that takes observables to states and vice versa.
By this we mean that if $A$ transforms as a state then $\tilde A$ transforms
as an observable, i.e.
    \begin{eqnarray}\label{eq:duality}
    \widetilde{(M^\dagger AM)}=M^{-1}\,\tilde A\,(M^{-1})^\dagger
    \end{eqnarray}
For a single qubit the tilde map is given by
    \be
    \tilde A= \sigma_3A^t\sigma_3
    \ee
and for a pair of qubits by
    \be
    \tilde A= (\sigma_3\otimes \sigma_3)\,A^t\,(\sigma_3\otimes \sigma_3)
    \ee
That the tilde map indeed satisfies Eq.~(\ref{eq:duality}) is a
property of $\sl$. It follows from the identity
    \begin{equation}\label{sl2}
    \sigma_3 M^t\sigma_3=M^{-1}, \quad  M\in\sl
    \end{equation}
The tilde operation acts on the Pauli spin matrices as
``spin-flip", reversing the spatial component \footnote{The
notation used in high energy physics \cite{peskin} is bar rather
then tilde. We use tilde to be consistent with
Wootters\cite{Wootters:concurrence}.},
    \be\label{eq:spin-flip}
    \tilde{\sigma}_\mu=\sigma_3 \sigma^t_\mu\sigma_3=\sigma^\mu
%%  =(\sigma_0,-\sigma_i)
    \ee
(Indices are raised and lowered according to the Minkowsky metric
tensor $\eta$). It follows that
    \be
    Tr\, (\sigma_\mu\tilde\sigma_\nu)= 2\eta_{\mu\nu}
    \ee

If we represent single qubit states and observables by
contravariant components of 4-vectors,
    \be
    \rho={\sf r}^\mu \sigma_\mu,\quad W={\sf w}^\mu \tilde\sigma_\mu
    \ee
the invariance of $Tr (\rho {W})$ implies that ${\sf r}^\mu$ and
${\sf w}^\mu$ transform under the same Lorentz transformation:
    \be\label{eq:rw}
    Tr (\rho {W}) =2\,{\sf r}^\mu{\sf w}^\nu\eta_{\mu\nu}
    \ee
This carries over to 2-qubits where states and witnesses are
represented by contravariant tensors
    \be\label{eq:A-lorentz-comp-1}
    \rho ={\sf r}^{\mu\nu}\sigma_\mu\otimes \sigma_\nu,\quad
    W={\sf W}^{\mu \nu}\tilde\sigma_\mu\otimes\tilde
    \sigma_\nu
    \ee
and now the Lorentz scalar is
 \be\label{eq:rw-tensor}
    Tr (\rho {W}) =4\,{\sf r}^{\mu\nu}{\sf W}_{\mu\nu}
    \ee

%%%%%%%%%%%%%%%%%%%%%%%%%%%%%%%%%%%%%%%%%%%%%

\subsection{Self-duals and Anti-self-duals}

We shall say that $W_{e}$ is self-dual if
    \begin{equation}\label{self-dual}
    W_{e}=\tilde W_{e}
    \end{equation}
For a single qubit self-duality means that the state is fully
mixed. For 2-qubits it readily follows from
Eq.~(\ref{eq:spin-flip}) that the matrix of Lorentz components of
$W_{e}$ has the form
    \be\label{eq:ae-self-dual}
    {\sf W}_{e}=\left(%
    \begin{array}{cccc}
    {\sf W}_{00} & 0 & 0 & 0 \\
    0 & {\sf W}_{11} & {\sf W}_{12} & {\sf W}_{13} \\
    0 & {\sf W}_{21} & {\sf W}_{22} & {\sf W}_{23} \\
    0 & {\sf W}_{31} & {\sf W}_{32} & {\sf W}_{33} \\
    \end{array}%
    \right)
    \ee
The space of self-duals is then evidently 10 dimensional.
Self-dual states represent fully mixed subsystems. Self-dual
observables are time-reversal invariant.

Since $SU(2)$ act on the (spatial components) of the Pauli
matrices like a rotation, the singular value decomposition implies
that

    \begin{prop}\label{prop:s-d-diag-form}
     Every self-dual $W_e$ can be brought to a canonical form
    \be
    W_e\to {\sf w}_\mu \sigma_\mu^{\otimes 2}=\omega_\mu P_\mu
    \ee
    by a pair of $SU(2)$ transformation, where ${\sf w_0}={\sf W}_{00}$ and
    the spatial components, ${\sf w}_j$ are the singular values of the $3\times 3$ sub-matrix of
     ${\sf W}_e$ (up to sign\footnote{Singular value decomposition requires $O(3)$ while we use only $SO(3)$.}).
The $\omega_\mu$ and ${\sf w}_\mu$  are related by the linear
transformation of Eq.~(\ref{eq:bell})
    \begin{eqnarray}\label{eq:invariant-spectrum-coordinates}
    \left(\begin{array}{c}
    \omega_0 \\
    \omega_1 \\
    \omega_2 \\
    \omega_3
    \end{array}\right)=\left(
                     \begin{array}{crrr}
                         1 &  1 &  1 & -1 \\
                         1 &  1 & -1 &  1 \\
                         1 & -1 &  1 &  1 \\
                         1 & -1 & -1 & -1 \\
                    \end{array}
                   \right)\left(\begin{array}{c}
    {\sf w}_0 \\
     {\sf w}_1 \\
    {\sf w}_2 \\
    {\sf w}_3
    \end{array}\right)
    \end{eqnarray}
In particular, the $\omega_\mu$ are all real.
    \end{prop}

The space of (not necessarily hermitian) anti-self-dual operators
may be identified with the Lie algebra of $\sl\otimes\sl$. Indeed
by Eq.~(\ref{sl2})
    \begin{equation}\label{mm}
    \tilde M=M^{-1},\quad \forall M\in \sl\otimes\sl
    \end{equation}
from which follows
    \be
    \delta\tilde M= -M^{-1}\, (\delta M) \, M^{-1}
    \ee
The Lie algebra is the variation at the identity where
$\delta\tilde M=-\delta M$, i.e.  the Lie algebra is
anti-self-dual. Since the linear space of anti-self-dual operators
is %easily seen to be
6-complex-dimensional (spanned by $\sigma_i\otimes1,\,
1\otimes\sigma_i;\; i=1,2,3$) it must coincide with the whole Lie
algebra.

\subsection{ LSL invariants}\label{sec:invariant-spectrum}

Since $\tilde A$ transform contragradiently to $A$, the product
$\tilde A A$ undergoes a similarity transformation by
Eq.~(\ref{eq:AW^diamond}). This allows us to associate spectral
invariants with the LSL action:
\begin{lemma}\label{lemma:invariants}
For any (n-qubit) observable, the spectrum of $\tilde A A$ and
$\det A$ are LSL invariants.
\end{lemma}
To get a feeling for the invariants consider first the case of a
single qubit state $\rho={\sf r}^\mu\sigma_\mu $. In this case
there is just a single invariant
    \be\label{eq:tilde pure}
     \tilde \rho \rho=  ({\sf r}_\mu {\sf r}^\mu)\, \one=(\det \rho)\, \one
    \ee
By Eq.~(\ref{eq:entropy}) the determinant encodes the information
on the entropy of the (normalized) state and vanishes for pure
states.

Multi-qubits observables of the form $\rho=\rho_A\otimes \rho_B$
then have as invariant $\det (\rho_A)\det(\rho_B)$. In particular,
it follows that
    \be\label{eq:alice-pure}
    \tilde\rho \rho=0
    \ee
     whenever either
$\rho_A$ or $\rho_B$ is a pure state.

In the case of two qubits, $spec(\tilde{A}A)$ gives four
LSL-invariants. The invariant $\det A$ is closely related to them,
since
    \be
    \det (\tilde A A)= (\det A)(\det \tilde A) =(\det A)^2
    \ee
Thus the only additional information supplied by $\det A$
is its sign.

In the particular case where $A$ is a state we trivially have $\det A\geq0$.
As $\tilde \rho \rho$ is readily seen to be
similar to the positive operator %%\cite{Wootters:concurrence}
    \be
    \sqrt \rho\, \tilde \rho \sqrt \rho,
    \ee
one evidently has in this case
    \begin{lemma}\label{lemma:wootters} (Wootters \cite{Wootters:concurrence})
    When $ \rho> 0$  the eigenvalues of $\tilde \rho \rho$ are all positive.
    \end{lemma}
We define the {\em LSL invariant spectrum  of state $\rho$} as the
the positive roots of the eigenvalues of $\tilde \rho \rho$.

Since a witness $W$ is, in general, not positive, it is not
a-priori clear that eigenvalues of $\tilde W W$ are positive.
In fact, for a general observable $A$, the spectrum of $\tilde A A$
need not even be real. We shall see, in the next section,
that for witnesses the eigenvalues of $\tilde W W$ are still
positive. Moreover, as we shall see, there is a natural way to
choose signs for the roots of these eigenvalues in a way that is consistent
with the invariance of $\det W$.
This will allow defining  the LSL invariant spectrum of $W$ in a way
that amalgamate the two invariants of  lemma \ref{lemma:invariants}.

%%%%*****************************

\section{Canonical forms as optimizers}

We want to extend the notion of LSL invariant spectrum to
witnesses. Since witnesses are, in general, not positive, the
argument leading to lemma \ref {lemma:wootters}  does not  apply.
However, as we shall see, the LSL invariant spectrum
$\{\omega_\mu\}$ is still well defined and real. In fact the
following key result holds:
    \begin{thm}\label{thm:can-form}~
    Any observable $W$ in the interior of the cone of potential
    witnesses is LSL equivalent to a witness of {\em canonical  form}
    \be\label{canonical}
    W\mapsto \omega_\mu P_\mu
    \ee
where $P_\mu$ are the Bell projections, $\omega_\mu\in\mathbb{R}$,
and $\omega_\mu^2$ are the eigenvalues of $\tilde W W$. The
representation (\ref{thm:can-form}) is unique, up to permutations
of the $\omega_\mu$'s. This generate the tetrahedral group
manifest in the Fig. \ref{fig:octa-tetra-cube}. A unique
representation is obtained by imposing the canonical order
    \be\label{ccoo}
    \omega_0\ge \omega_1\ge\omega_2\ge |\omega_3|
    \ee
In particular, at most one LSL-eigenvalue, the one with smallest
magnitude, is negative.
    \end{thm}
The upshot of this theorem is that the LSL equivalence classes of
the 16 dimensional cone of potential witnesses, and therefore also
the cone of (un-normalized) states, can be represented by points
in $\mathbb{R}^4$.

The proof of this theorem depends on a variational principle.
Specifically on finding a witness that maximizes the entanglement
evidence, in the sense of definition \ref{def:evidence}. This
point of view leads to our second key result:
    \begin{thm}\label{thm:optimal-witness}
    Let $W$ and $\rho$  be in the interior of the cone of potential
    witnesses, and let $W_e=\omega_\mu P_\mu$ and $\rho_e=\rho_\mu P_\mu$ be their
    associated canonical forms. Then
    \be
    \min_M\left\{Tr ( W^M\rho)\right\} =Tr(W_e\rho_e)= \sum_\mu \rho_\mu\omega_\mu
    \ee
where $\rho_\mu$ are chosen in canonical order, Eq.~(\ref{ccoo}),
while $\omega_\mu$ are chosen with the {\em anti-canonical order}
    \be\label{aco}
   \omega_0\le\omega_1\le \omega_2\le\omega_3
    \ee
In particular, taking $ \rho=1$, we see that the LSL map $W\mapsto
\omega_\mu P_\mu$ is trace decreasing.
\end{thm}
\begin{cor}\label{cor:witnesses-suffice}
It follows that to test whether a state is entangled it is enough
to test its canonical representer against canonical witnesses.
%%  $\omega_\mu P_\mu$.
\end{cor}
The proofs of both theorems are given in the following
subsections.

\subsection{Existence of the optimizer}
Consider the stationary points of the function $M\mapsto Tr(\rho
W^M)$ where $M=M_A\otimes M_B$.  For $M_{A,B}\in SU(2)$, this
function must have (finite) maximum and minimum\footnote{In fact
by Morse theory it must have at least one maximum, one minimum and
two saddles.} since $SU(2)$ is compact. However, in the case that
$M_{A,B}\in \sl$, which is not compact, there may be no stationary
point for any finite $M$. The existence of a minimum is guaranteed
by the following lemma.

  \begin{lemma}\label{lemma:oded}
    %\begin{itemize}
    Suppose $\rho$ and $W$ both lie in the interior of the cone of
    potential witnesses, i.e. satisfying a strict inequality in (\ref{eq:witness}).
    Then, the function  $Tr\left(\rho W^{M}\right)$, diverges to $+\infty$
    as either $M_{A}\in\sl$ or $M_B\in\sl $  go to infinity. In particular,
    it has a finite minimizer.
    \end{lemma}
%%%%%
The lemma may be written in a more symmetric  form  (under
$\rho\leftrightarrow W$) by noting
    \be\label{eq:variation of both}
    Tr(\rho^M W^N)= Tr(\rho  W^{(M^{-1})^\dagger N}).
    \ee

Sketch of the proof: The spectrum of any $M_A\in\sl$
%element
is of the form $\{\lambda,1/\lambda\}$. The element is large when
$|\lambda|$ is large. It can then be approximated by a rank one
operator (corresponding to the large eigenvalue) $M_A\simeq
\lambda P$, with $P$ a one dimensional projection. Thus
$M=M_A\otimes M_B\simeq\lambda P\otimes M_B$ is essentially
supported on a $1\otimes2$ dimensional subspace of the full
$2\otimes2$ space. As a $1\otimes2$ space cannot support any
entanglement, the corresponding expectation $Tr\left(\rho
W^{M}\right)\simeq|\lambda|^2Tr\left(\rho W^{P\otimes M_B}\right)$
must be positive. As it is multiplied by $|\lambda|^2$ it actually
diverge to $+\infty$ with $\lambda$.  \hfill$\square$

An alternative proof of lemma \ref{lemma:oded} and a
generalization of it which applies to witness on the boundary are
described in appendix \ref{sec:main}.

\subsection{Characterization of the optimizer}
In the previous section we have seen that $Tr (\rho W^M)$ has a minimizer.
Once its existance is guaranteed, one may use standard
variational procedure to characterize it.
As we shall show whenever $\rho$ is of the canonical form
the minimizer---the optimal witness,
is also of the form $\omega_\mu P_\mu$.

Note that the Bell projections are self-dual, $P_\mu=\tilde
P_\mu$. We start by showing that the stationary points of $Tr
(\rho W^M)$ are self-dual.
    \begin{lemma}\label{lem:stationary}
       The stationary points of the function $Tr (\rho W^M)$,
       where $M=M_A\otimes M_B$ with
    $M_{A,B}\in \sl$ are the self-dual points, i.e.
    \be \label{eq:stationary}
    \widetilde{\rho W^M}=\rho W^M
    \ee
    \end{lemma}
%%%%%%%%%%%%%%%%%%%%
%%%%%%%%%%%%%%%%%%%%%%%%%%%%
Proof:
Suppose  $Tr (\rho W^M)$ is stationary at the identity $M=I$
then for any small LSL-variation $M=I+\delta M$ we must have
    \bea
    0&=&\delta Tr (\rho W^M)=\delta Tr(\rho M^{-1}W{M^{-1}}^\dagger)\nonumber \\
    &=&
    Tr\left(\rho (-\delta M)W+\rho W(-\delta M^\dagger)\right) \nonumber \\
    &=&-Tr(W\rho\delta M)-Tr(W\rho\delta M)^*\\ &=&-2Re\,Tr(W\rho\delta
    M)\nonumber
    \eea
where we used the fact that $\rho,W$ are hermitian.
Recall that by Eq.~(\ref{mm}) the Lie algebra of LSL
consists of complex matrices satisfying $\tilde{\delta M}=-\delta M$.
Stationarity requires $W\rho$ to be in the space orthogonal to these
which is the space of self-duals $ W \rho=\widetilde{W \rho}$.
Formally this follows by using the identity $Tr(\tilde{A}\tilde{B})=Tr(AB)$
to write
    \bea
    0=\delta Tr (\rho W^M)
    =- Re\,Tr \bigg(\delta M\, \big( W \rho-\widetilde{W    \rho}\big)\bigg)
    \eea
As both $\delta M$ and  $ W \rho-\widetilde{W \rho}$ are
anti-self dual, the trace of their product vanishes for arbitrary
(complex) anti self-dual $\delta M$ if and only if  $ W \rho=\widetilde{W \rho}$.
A stationary point at arbitrary $M$ similarly lead to (\ref{eq:stationary}). \hfill $\square$
%%%%%%%%%%%%%%%%%%%%%%%%%%%%%%

It follows that any strict potential witness $W$ is LSL-equivalent to
a self-dual one. To see this take $\rho=1$.
Lemma \ref{lemma:oded} then guarantees that $Tr({W}^M)$ has a
minimum, and lemma \ref{lem:stationary} tells us that the
minimizer ${W}^M$ is self dual.
Moreover, by applying Prop. \ref{prop:s-d-diag-form} it follows
that $W$ can be brought to a canonical form, Eq.~(\ref{canonical}).
The lemma below gives a direct proof
of this fact without relying on the singular value decomposition
used in Prop. \ref{prop:s-d-diag-form}.
    \begin{lemma}\label{nhv}
    Suppose  the state $\rho$ is self-dual, then $Tr(\rho  W^M)$ has
    its stationary points where $[\rho^2, W^M]=0$. In particular if $\rho$ is
    in canonical form Eq.~(\ref{canonical}),
    then (at least in the generic non-degenerate case) so is the minimizer ${W}^M$.
    In the case of degenerate $\rho^2$ the minimizer ${W}^M$ is
    not unique but may still be chosen as canonical.
    \end{lemma}
Proof: Combining $\rho=\tilde{\rho}$ with the minimizer condition
(\ref{eq:stationary}) gives
    \be
    (\rho W^M)=\widetilde{(\rho W^M)}= \widetilde{(W^M)}\tilde{\rho}=
    \widetilde{(W^M)}\rho
    \ee
and similarly
    \be
    (W^M\rho)=\widetilde{(W^M\rho)}=\tilde{\rho}\widetilde{(W^M)}=\rho\widetilde{(W^M)}
    \ee
Combining the two  gives $\rho^2 W^M=W^M\rho^2$.
%% Thus $W^M,\rho^2$ may be simultaneously diagonalized.
Thus the Bell basis which diagonalize $\rho$ must do the same for $W^M$,
unless $\rho^2$ happens to be degenerate.
In the special case of degenerate $\rho^2$ one may find a
Bell-diagonalized minimizer $W^M$ by considering first a small
degeneracy breaking perturbation of $\rho$. \hfill$\square$

\subsection{Proofs of the two theorems}

Proof of  theorem \ref{thm:can-form}: Choosing some generic $\rho$
of the canonical form (\ref{canonical}) and arbitrary $ W$, we are
guaranteed by lemma \ref{lemma:oded} that $Tr(\rho{W}^M)$ has a
minimum. Lemma \ref{nhv} then tells us that the minimizer $W^M$ is
also of the canonical form $\omega_\mu P_\mu$.
The four eigenvalues $\omega_\mu$ can
be permuted arbitrarily by Proposition~\ref{prop:permutation}.

The LSL invariance of
    \be
    Spec(\tilde{W}W)=\{\omega_\mu^2\}
    \ee
determines $\omega_\mu$ up to sign. To determine the signs we shall
show that at most one LSL-eigenvalue is negative. To this end, note
first that
from any pair of the Bell states, one may consruct a  separable state
$\ket{\beta_\mu}+e^{i\varphi}\ket{\beta_\nu}$ (actually $e^{i\varphi}=1$ or $i$).
E.g.
    \be
    \ket{\beta_0}+ \ket{\beta_1}=\sqrt 2\ket{00}
    ,\;\ket{\beta_2}+ i\ket{\beta_3}=\sqrt 2\ket{01}, etc
    \ee
(Since local unitaries can permute Bell states, similar relations must hold for
all other Bell pairs.)
With $W$ in canonical form we then have by Eq.~(\ref{eq:witness}) that
    \be
    0\le\bra{\beta_\mu+e^{-i\varphi}\beta_\nu}W^M\ket{\beta_\mu+e^{i\varphi}\beta_\nu}=
    (\omega_\mu+\omega_\nu),\quad \mu \neq \nu
    \ee
This imply that at most one of the eigenvalues  is negative, and
moreover, it must be the one of smallest absolute value. The
LSL-invariance of $\det(W)$ then fixes the signs of all the
$\omega_\mu$ uniquely. The proof is complete. \hfill$\square$

Proof of theorem \ref{thm:optimal-witness}: Given $W$ and $\rho$
theorem \ref{thm:can-form} tells us that that
$\rho^N=\rho_e=\rho_\mu P_\mu$ for some $N\in\sl\otimes\sl$. We
therefore have

    \be
    Tr(\rho{W})=Tr(\rho_e^{N^{-1}}{W})=Tr(\rho_e{W}^N)\geq
    \min_M\left\{Tr(\rho_e{W}^{M})\right\}
    \ee
By lemma \ref{nhv} we know that $Tr(\rho_e{W}^{M})$ is stationary
whenever ${W}^M= \omega_\mu P_\mu$. The minimum clearly
corresponds to requiring
Eqs.~(\ref{ccoo},\ref{aco}).\hfill$\square$

%%%%%%%%%%%%%%%%%%%%%%%%%%%%%%%%%%%%%%%

\subsection{The boundary of the cone of witnesses}
Theorem \ref{thm:can-form} applies only to observables $W$ in the
interior of the cone of potential witnesses. It can be extended to
to the boundary, provided the notion of LSL-transformations is
appropriately modified. We shall say that observable $B$ is
obtained by a generalized LSL-transformation from observable $A$
iff it is in the closure of the equivalence class of $A$
\footnote{This is not an equivalence relation as generalized
transformations need not be invertible. }, i.e. iff there exist a
series $\{M_i\}_{i=1}^\infty\subset\sl\otimes\sl$ such that
$A^{M_i}\rightarrow B$. Theorems
\ref{thm:can-form},\ref{thm:optimal-witness} then hold for any
potential witness  with $\min_M$ replaced by $\inf_M$. The proof
follows very similar lines to the proofs given above. The only
major change is replacing lemma \ref{lemma:oded} by a
generalization of it described in the appendix \ref{sec:main}.

%%%%%%%%%%%%%%%%%%%%%%%%%%%
%%%%%%%%%%%%%%%%%%%%%%%%%%%%%
\section{Classification: A Lorentzian picture}\label{sec:potential-witnesses}

\subsection{Geometric characterization of witnesses}

The matrix of Lorentz components ${\sf W}_{\mu\nu}$ of a potential
witness $W$, Eq.~(\ref{eq:A-lorentz-comp-1}), allows for a simple
geometric characterization of potential witnesses. By definition,
a potential witness has positive expectation for product states,
\begin{equation}\label{eq:positive-lc}
0\le Tr \, ({ W} { \rho}_A\otimes { \rho}_B)=4\,{\sf W}_{\mu\nu}
\, (\rho_A)^\mu (\rho_B)^\nu .
\end{equation}
Since the 4-vectors $\left(\rho_{A,B}\right)_\mu$ can lie anywhere
in the forward light cone (by lemma \ref{lem:light-cone}), we
learn that the matrix ${\sf W}$ maps the forward light cone
into itself.  Points that lie in the interior of the cone of
potential witnesses satisfy a strict inequality in
Eq.~(\ref{eq:positive-lc}). The map ${\sf W}$ then sends the
forward light cone into its (timelike) {\em interior}.

\subsection{Lorentz singular values}\label{sec:lsv}

By Eq.~(\ref{lorentz})  LSL acts on the Lorentz components of a
2-qubit observable $W$ by a pair of Lorentz transformations
\begin{equation}\label{lorenta}
{\sf W}^{A\otimes B}= \Lambda_{A} \,{\sf W}\, \Lambda_B^t
\end{equation}
where $\Lambda_{A,B}\in SO_+(1,3)$ are two $4\times 4$ Lorentz
transformation matrices.

From Eq.~(\ref{eq:bell}) it follows that if $W$ is in canonical
form then
    \be\label{eq:2can-forms}
    W=\omega_\mu P_\mu= {\sf w}_\mu \sigma_\mu^{\otimes 2}
    \ee
so that the Lorentz matrix ${\sf W}$ is diagonal.  The
$\omega_\mu$ and ${\sf w}_\mu$ coordinates are related by
Eq.~(\ref{eq:invariant-spectrum-coordinates}).
%
%    \begin{eqnarray}\label{eq:invariant-spectrum-coordinates}
%    \left(\begin{array}{c}
%    \omega_0 \\
%    \omega_1 \\
%    \omega_2 \\
%    \omega_3
%    \end{array}\right)=\left(
%                     \begin{array}{crrr}
%                         1 &  1 &  1 & -1 \\
%                         1 &  1 & -1 &  1 \\
%                         1 & -1 &  1 &  1 \\
%                         1 & -1 & -1 & -1 \\
%                    \end{array}
%                   \right)\left(\begin{array}{c}
%    {\sf w}_0 \\
%     {\sf w}_1 \\
%    {\sf w}_2 \\
%    {\sf w}_3
%    \end{array}\right)
%    \end{eqnarray}
%

Thus when viewed in terms of Lorentzian components, bringing $W$
to its canonical form consists of diagonalizing the associated
tensor ${\sf W}_{\mu\nu}$ by a pair of Lorentz transformations.
This is reminiscent of the  notion of the singular decomposition
of a matrix (which is defined in the same way with the Lorentz
transformation replaced by orthogonal matrices).

For a matrix $M$ its singular values are the (positive) roots of
the matrix $M^\dagger M$. The Lorentzian analog of $M^\dagger M$
turns out to be the  matrix ${\sf W}^{\lambda\mu}{\sf
W}_{\lambda\nu}$.
%%It spectrum is LSL invariant and is an analog of section \ref{sec:invariant-spectrum}.To see this,%%
It is convenient to write it as ${\sf W}^\star{\sf W}$ where the
`Lorentz conjugated matrix' is defined by\footnote{The $\star$
duality,  conforms with the Lorentz scalar product,
    $
   v \cdot({\sf W} u)    =({\sf W}^{\star }v)\cdot u
   $. It is distinct from the $\tilde{} $
duality of the previous section. }
\begin{equation}\label{lorentz-conjugate}
{\sf W}^\star=\eta {\sf W}^t \eta
\end{equation}
One readily verifies that componentwise
    \begin{equation}\label{eq:A^star-munu}
    ({\sf W}^\star)_{\mu\nu}={\sf W}^{\nu\mu}
    \end{equation}
%In particular the matrix $W_{\mu}^{\:\nu}$ is $\star$ self dual
%iff the tensor $W_{\mu\nu}$ is symmetric.
Since Lorentz transformations leave the Minkowsky metric
invariant, $\Lambda^t \eta \Lambda =\eta$, one has
\begin{equation}\label{lorentz-define}
\Lambda^{\star } =\Lambda^{-1}
\end{equation}
It then follows that under (\ref{lorenta})
${\sf W}^\star {\sf W}$ undergoes a similarity transformation
\begin{equation}\label{eq:a-star-a}
{\sf W}^\star {\sf W}\to \big(\Lambda_M{\sf
W}\Lambda_N^t\big)^\star \big(\Lambda_M{\sf W}\Lambda_N^t\big)=
(\Lambda_N^t)^{-1}\big({\sf W}^\star {\sf W}\big)\Lambda_N^t
\end{equation}
and similarly for ${\sf W} {\sf W}^\star$. The spectra of ${\sf
W}^\star {\sf W}$ and ${\sf W} {\sf W}^\star$ are therefore LSL
invariant.

The LSL invariance of the spectrum of ${\sf W} {\sf W}^\star$ does
not depend on $W$ being a witness. In general, this spectrum is
complex. For matrices that lie in the  cone of
witnesses one has, by Eq.~(\ref{eq:2can-forms}), that the eigenvalues
of ${\sf W} {\sf W}^\star$ are ${\sf w}_\mu^2$ and are positive (or zero).
In this case, in analogy with the notion of singular values, one
may define the Lorentz singular values as $|{\sf w}_\mu|$.

\begin{rem}
%Indeed using
Diagonal Lorentz transformations in $O(1,3)$ with $\pm1$ on the
diagonal, bring any diagonal $W$ to a positive diagonal form.
However, we are allowed only proper orthochronos Lorentz
transformations, $ SO_+(1,3)$. Thus the canonical coordinates
${\sf w}_\mu$ defined via Eq.~ (\ref{eq:2can-forms}) may differ in
signs from the (positive) Lorentz singular values.
%% Orthochronality implies a fixed sign of ${\sf w}_0$ the
%% Lorentz singular value associated with a time like direction.
\end{rem}

\subsection{Tetrahedral symmetry and fundamental domains}

The tetrahedral group acts on the coordinates $\omega_\mu$ as
permutations. In terms of the coordinates ${\sf w}_\mu$ this group
acts as permutations and sign flips of the three `spatial'
coordinates ${\sf w}_j$ which leave
$sgn({\sf w}_1{\sf w}_2{\sf w}_3)$  invariant.
To see this note first that the relation $ 4{\sf w}_0=
\omega_0+\omega_1+\omega_2+\omega_3>0 $ shows that
${\sf w}_0$ is independent of the  ordering of $\omega_\mu$.
%% (which is consistent with orthochronality.) %%HHHHH
Hence,  the tetrahedral group acts only on the spatial components ${\sf w}_j$.
For proper Lorentz transformations
$\det({\sf W})={\sf w}_0{\sf w}_1{\sf w}_2{\sf w}_3$ cannot change
sign. In cases when $\det({\sf W})>0$ the canonical coordinates
may be taken as equal to the Lorentz singular values. If
$\det({\sf W})<0$ then at least one of the canonical coordinates
(which we will usually take to be the one having least absolute
value) must be chosen as negative. %%
    %\item
The tetrahedral symmetry allows one to impose ${\sf w}_\mu$ to be
in the {\em fundamental domain}
  \be\label{eq:fundamental-domain}
     {\sf w}_1\ge{\sf w}_2\ge |{\sf w}_3|
    \ee
which is equivalent to Eq.~(\ref{ccoo}). The
     {\em antipodal fundamental domain}
     \be\label{eq:anti-fundamental-domain}
     -{\sf w}_1\ge-{\sf w}_2\ge |{\sf w}_3|
     \ee
is equivalent to the anti-canonical ordering of Eq.~(\ref{aco}).

%%%%%

%%%%%%%%%%%%%%%%GGGGGGGGGGGGGGGGGGGGGGG*\marginpar{problems.}

%%%%%%%%%%%%%%%%%%%% %%%%%%%%%%%%%%%%%%%%%%%%%%

\subsection{Classification of potential witnesses}

%In this section we show that the LSL equivalence classes of the
%the 16 dimensional cone of potential witnesses make
%a cone in 4 dimensions. Points on the boundary of this 4
%dimensional cone can describe more than one LSL equivalence
%class (which we also completely characterize). The basic idea
%of the construction is to choose an appropriate pair of Lorentz
%transformations that will bring the matrix of Lorentz coefficients
%${\sf W}$ of a potential witness $W$ into a ``canonical" diagonal
%form. This is not possible for a general matrix, but is possible
%for {\em almost all} potential witnesses.
%\footnote{Allowing infinite boosts would enable diagonalizing all witnesses.}

Symmetric matrices that map the forward light-cone into itself may
be interpreted in general relativity as energy-momentum tensors
that satisfy the ``dominant energy condition''. Their
classification\footnote{Landau and Lifshitz, p. 274 in
\cite{Landau-Lifshitz}, gives a partial classification.} is given
in p. 89-90 in \cite{hawking}.  We need a generalization of this
classification to non-symmetric matrices
\footnote{By(\ref{lorentz-conjugate}) $W^\star=W$ means $W\eta$ is
symmetric.}
 where we are allowed to use a
pair of Lorentz transformations as in Eq.~(\ref{lorenta}). The
classification is given in \cite{Verstraete:local-filtering} and
is based on \cite{Gohberg}:
%\marginpar{uniqueness ? with\ref{eq:fundamental-domain}}
    \begin{thm}\label{thm:classification}
    Let ${\sf W}$ be a $4\times 4$  matrix that maps the forward
    light-cone into itself. Fix arbitrary $\kappa>0$. Then there is
    a pair of Lorentz transformations $\Lambda_A,\Lambda_B$ such that
    $\Lambda_A{\sf W}\Lambda_B$ is of one of the 4 canonical
    forms, unique subject to Eq.~\ref{eq:fundamental-domain}:
        \begin{itemize}
        \item The ordinary diagonal form
        \begin{equation}\label{eq:diagonal-can-form}
        \left(
      \begin{array}{cccc}
        {\sf w}_0 & 0 & 0 & 0 \\
        0 & {\sf w}_1 & 0 & 0 \\
        0 & 0 & {\sf w}_2 & 0 \\
        0 & 0 & 0 & {\sf w}_3 \\
      \end{array}
    \right),
    \end{equation}
    associated with the cone in 4 dimensions with a cross section
    that is a 3 dimensional cube:
    \be
    \quad {\sf w}_0\ge |{\sf w}_j|
    \ee
    \item The first extraordinary form
    \begin{equation}\label{eq:except-can-form1}
    \left(
      \begin{array}{cccc}
        {\sf w}_0+\kappa & -\kappa & 0 & 0 \\
        \kappa  & {\sf w}_1-\kappa & 0 & 0 \\
        0 & 0 & {\sf w}_2 & 0 \\
        0 & 0 & 0 & {\sf w}_3 \\
      \end{array}
    \right),
    \end{equation}
    associated with the boundary  of the cone
        \be
        {\sf w}_0={\sf w}_1\ge  |{\sf w}_{2,3}|
        \ee
\item The second extraordinary form
\begin{equation}\label{eq:except-can-form2}
    \left(
      \begin{array}{cccc}
        \kappa & \kappa & 0 & 0 \\
        0 & 0 & 0 & 0 \\
        0 & 0 & 0 & 0 \\
        0 & 0 & 0 & 0 \\
      \end{array}
    \right)
    \end{equation}
    associated with the apex of the cone ${\sf w}_\mu=0$.
    \item The third extraordinary form
    \begin{equation}\label{eq:except-can-form3}
    \left(
      \begin{array}{cccc}
        \kappa & 0 & 0 & 0 \\
        \kappa & 0 & 0 & 0 \\
        0 & 0 & 0 & 0 \\
        0 & 0 & 0 & 0 \\
      \end{array}
    \right),
\end{equation}
also associated with the apex of the cone ${\sf w}_\mu=0$.
\end{itemize}
${\sf w}_\mu$, the ``Lorentz singular values'', are roots of the
eigenvalues of $W^\star W$.
\end{thm}

%%%%%%%%%%%%%%%%%%%%
The proof of this theorem is given in appendix \ref{appendix:lsv}.

%This classification allows us to describe both the ordinary and
%extraordinary witnesses.
%and states in the world of two qubits
%geometrically in 4 dimensions.

%%%%%%%%%%%%%%%%%%%%%%%%%%%%%%%%%%%%%%%%%%%%%%%%%%%%%%%%%%%%%%%%%%%%%%

\section{The Geometry of witnesses and states}
We have seen that the LSL equivalence classes of
witnesses, states and separable states are represented by nested
cones in four dimensions.
In this section we give a geometric description of these cones.

%%%%%%%%%%%%%%%%%%%%%%%%

\subsection{The geometry of ordinary witnesses}\label{sec:geometry-witnesses}

%%%%%%%%%%%%%%%%%%%%%%%%%%%%%%%%%%%%%%%%%%%%%%%%%%%%%%%

%%************
A diagonal witness
    \be
    W_e={\sf w}_{\mu} \sigma_\mu^{\otimes 2}
    \ee
maps the light cone into itself iff ${\sf w}_0\ge |{\sf w}_j|$.
The LSL equivalence classes of the (ordinary) potential witnesses
are therefore characterized geometrically by the cone in 4
dimensions:
                \begin{equation}\label{eq:cube}
                 {\sf w}_0\ge |{\sf w}_1|,|{\sf w}_2|,|{\sf w}_3|
                \end{equation}
whose cross section is the cube.

By theorem  \ref{thm:optimal-witness} the canonical representative
of a witness also minimizes $Tr(W^M)$. Thus the representatives of
normalized witnesses have
$
%    \be
    {\sf w}_0 \le \quarter
%   \ee
$
giving a capped cone. All points in the capped cone are relevant
since given $W_e\neq 0$ with $Tr\ W_e<1$ one easily finds $M$
which makes $Tr(W_e^M)$ as large as one wants.

Four corners of the cube at the cap of the cone, making the
vertices of a tetrahedron, represent the 4 Bell states $P_\mu$.
The four remaining corners, also making a tetrahedron, describe
bona-fide Bell witnesses, all equivalent to the swap
$S=\half\sum\sigma_\mu^{\otimes2}$.
   %% which lies at the corner of the anti-canonical fundamental domain given by
   %% \be
   %%  1>\sum\omega_\mu,\;\;\;\;
   %%  |\omega_0|\le\omega_1\le\omega_2\le\omega_3
   %%  \ee

\subsection{The geometry of the ordinary separable states}
The duality between separable states and potential witnesses in 16
dimensions translates to a duality between the cones of the
corresponding equivalence classes in 4 dimensions. This follows
from corollary \ref{cor:witnesses-suffice} which says that the
corresponding cones in $\mathbb{R}^4$, defined by
$\rho_\mu\omega_\mu\geq0$ are also dual cones. The identity
$\rho_\mu\omega_\mu=4{\sf r}_\mu {\sf w}_\mu$ allows writing this
in terms of canonical coordinates as ${\sf r}_\mu {\sf
w}_\mu\geq0$. Since the dual of the cube is the octahedron, the
LSL equivalence classes of the separable states are represented by
a cone whose cross section is an octahedron.

Algebraically, the separable states are described  by the 8
extremal inequalities
    \be\label{...}
    \quarter \ge {\sf r}_0\ge{\sf w}_1{\sf r}_1+{\sf w}_2{\sf r}_2+{\sf w}_3{\sf r}_3
    , \quad {\sf w}_i=\pm1
    \ee
associated with the eight witnesses at the corners of the cube,
making up an octahedral cone.
%%\footnote{The set of all (ordinary) states on the other hand is described
%%by the 4 inequalities where $\sf w_1w_2w_3=1$.}

A different way \cite{Leinaas-Myrheim-Ovrum} to see that the
separable states are represented by the octahedron relies on
considering explicitly the 6 operators corresponding to the
vertices of the octahedron:
\begin{equation}\label{corners-separable}
8S_{j\pm}=(\sigma_0+\sigma_j)\otimes
(\sigma_0\pm\sigma_j)+(\sigma_0-\sigma_j)\otimes
(\sigma_0\mp\sigma_j)=2(\sigma^{\otimes 2}_0\pm \sigma^{\otimes
2}_j)
\end{equation}
and $j=1,2,3$. The middle expression shows that all 6 vertices are
separable states. The right hand side shows that they all are
equal mixtures of any two Bell states.

\subsection{The geometry of all ordinary states}\label{sec:geomet-state}

%%%%%%%%%%%%%%%%%%%%%%%%%%%%%%%%%%%%%%%%%%%%%%%%%%%%%%%%%%%%%%%%%%%%%%
Let $\rho_e$ be a canonical representer corresponding to the state
$\rho$, i.e.
    \be
    \rho_e={\sf r}_\mu\sigma^{\otimes 2}_\mu=\rho_\mu
    P_\mu
    \ee
Since $\rho\ge 0$,  the LSL equivalence classes are represented by
the positive quadrant, $\rho_\mu\ge 0$, in 4 dimensions. This is
evidently a cone whose cross section is the tetrahedron.

In terms of the ${\sf r}$ coordinates the cone of all states is
described by 4 out of the 8 inequalities Eq.~(\ref{...}),
specifically those corresponding to ${\sf w}_1{\sf w}_2{\sf
w}_3=-1$.

The LSL equivalence classes corresponding to normalized states
form a 4 dimensional capped cone with
    \be\label{eq:half-space}
    \sum \rho_\mu\le 1
    \ee
The cap of the cone is the three dimensional tetrahedron,
%%A (generic) point on the cap represents a 6 dimensional $SU(2)$
%%equivalence class of states with fully mixed subsystems.
and represents the $SU(2)$ equivalence classes of states with fully
mixed subsystems.

The 4 vertices of the tetrahedron at the cap of the cone are
identified with the 4 Bell states $P_\mu$ of Eq.~(\ref{eq:bell})
and represent a single equivalence class as the tetrahedral
symmetry can interchange any of them, by Proposition
\ref{prop:permutation}. The $\rho_\mu$ coordinate lines represent
the (equivalence classes) of entangled pure states discussed in
section \ref{sec:pure-states}.

The apex of the cone at the origin formally corresponds to the
states where $\tilde \rho \rho=0$ which, by
Eq.~(\ref{eq:alice-pure}), occurs when at least one of the
subsystems is pure, as in Eq.~(\ref{mixed-pure}).

Any point in the cone of states can be expressed as a (sub) convex
combination of its vertices representing the four Bell
states\footnote{Using Eq. (\ref{extraordinary-bell}) one may
demonstrate the correctness of the corollary also for extraordinary
states.}
\begin{cor}
Any mixed 2 qubit state can be expressed as a convex combination
of 4 pure states, each equivalent to a Bell state by the same LSL-transformation.
\end{cor}

%%%%%%%%%%%%%%%%%%%%%%%%%%%%%%%%%%%%%%%%%%%%%%%%%%%%%%%%%%

%%%%%%%%%%%%%%%%%%%%%%%%%%%
%%%%%%%%%%%%%%%%%%%%%%%%%%%%
%%%%%%%%%%%%%%%%%%%%%%%%%%%%%%%%%%%%%%%%%%%%%%%%%%%%%%%%%%%%%%%%%%%%%%

The fundamental domain of normalized states is most simply described in terms
of its spectral coordinates $\rho_\mu$ as
$$\rho_0\geq\rho_1\geq\rho_2\geq\rho_3\geq0,\;\;\; \sum\rho_\mu\leq1$$
%Since we described the witnesses in terms of the canonical
%coordinates, it is useful to translate the states description
%given above using the spectral coordinates into the canonical
%coordinates. This is done via the linear transformation
%(\ref{eq:invariant-spectrum-coordinates}).
%%The LSL equivalence classes of normalized states are then uniquely
%%represented by the intersection of the fundamental domain
%%Eq.~(\ref{eq:fundamental-domain}) with the plane
%%    \be
%%    \quarter \ge {\sf r}_0\ge{\sf r}_1+{\sf r}_2+{\sf r}_3
%%    \ee
%%
%The LSL equivalence classes of normalized states are then represented as
%the fundamental domain
or, equivalently, by
    \be\label{eq:fundamental-domain1}
    \quarter \ge {\sf r}_0\ge{\sf r}_1+{\sf r}_2+{\sf r}_3,
    \;\;\;\;\;\;
    {\sf r}_1\geq{\sf r}_2\geq|{\sf r}_3|.
    \ee

%%%%%%%%%%%%%%%%%%%%%%%%%%%%%%%%%%%%%%%%%%

%%%%%%%%%%%%%%%%%%%%%%%%%%%%%%%%%ffffffffffffffffff

%%%%%%%%%%%%%%%%%%%%%%
%%%%%%%%%%%%%%%%%%%sec 7
%%%%%%%%%%%%%%%%%%%%%%%%

%%%%%%%%%%%%%%%%%%%% %%%%%%%%%%%%%%%%%%%%%%%%%%
\subsection{The geometry of the boundary}\label{sec:extraordinary}

The boundary of the cone of potential witnesses is subtle.
Observables inside the cone are guaranteed to have a finite LSL
transformation that brings them to canonical form. However, as one
approaches the boundary, it may happen that the required LSL
transformation may or may not have a limit. If it does, the
state/witness belongs to an ordinary class, if it does not, it
belongs to an extraordinary LSL equivalence class. Both classes,
though LSL inequivalent,  have identical invariant spectra and
Lorentz singular values and therefore are represented by the same
point in 4 dimensions. This makes the set of LSL equivalence
classes non-Hausdorff.

\begin{figure}[ht]
\hskip 4 cm
\includegraphics[width=6cm]{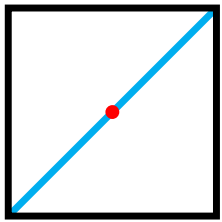}
\caption{The figure represents the extraordinary canonical forms
associated with the boundary of the cube,
Eq.~(\ref{extraordinary-bell}) with $\kappa\neq 0$. The
extraordinary separable states are represented by the red dot and
the extraordinary entangled states by the cyan diagonal. The
square represents extraordinary witnesses. The extraordinary
canonical forms are inequivalent to the ordinary ones
(corresponding to $\kappa=0$) cohabiting the same set of points.
} \label{fig:boudary}
\end{figure}

The first extraordinary family, Eq.~(\ref{eq:except-can-form1}) of
theorem \ref{thm:classification}, with ${\sf w}_1={\sf w}_0$ and
${\sf w}_2+{\sf w}_3\neq0$ describes observables with a negative
eigenvalue which therefore are witnesses rather than states. When
${\sf w}_2= -{\sf w}_3$ it describes the extraordinary family of a
mixture of two Bell states and a pure product state
    \begin{equation}\label{extraordinary-bell}
    \Big(p_0P_0+p_1P_1\Big)
    +{\kappa}(\sigma_0+\sigma_1)\otimes (\sigma_0-\sigma_1)
    \end{equation}
$p_0,p_1$  probabilities: $p_0+p_1\leq1$ and $p_0,p_1\geq0$.
For definiteness one may fix e.g. $\kappa=1$.
The Lorentz singular values are seen to be
    \be
    \frac {1} 4\{p_0+p_1,p_0+p_1,p_0-p_1,p_1-p_0\}
    \ee
Geometrically, this family
shown in Fig.\ref{fig:boudary}
may be thought of as
a phantom image of the edges of the  tetrahedron.

The second and third extraordinary forms describe the family
$(pure)\otimes(mixed)$ and $(mixed)\otimes(pure)$, both of which
are represented by the apex of the cone.
 %   \item

\section{Measure and distillation of entanglement}
The 4 dimensional description of the LSL equivalence classes of 2
qubits is faithful to the measure of entanglement. (This is not
true for the 3 dimensional description in
\cite{Leinaas-Myrheim-Ovrum}.) This allows us to give a geometric
interpretation of the notion of concurrence and optimize
distillation.
\subsection{Concurrence as the best evidence}

%%%%%%%%%%%%%%%%%%%%%%%%%%%%

A natural way to quantify entanglement is to measure the expected
values of entanglement witnesses \cite{Lewenstein}. Given an
entangled state $\rho$, the entanglement {\em evidence} given by
the expectation of the optimal  witness is
    \bea\label{ev}
    \textsf{C}(\rho)&=& -\inf_W 2\, Tr ( W \rho)
    =-8\,\inf_{{\sf w}} {\sf w}^\mu {\sf r}_\mu \nonumber \\
    &=& 2( - {\sf r}_0+ |{\sf r}_1|+ |{\sf r}_2|+ |{\sf r}_3|)%_+
    \eea
The set $W$ in this definition is the set of witnesses with a {\em normalized
representer}.
For a separable state the r.h.s. of Eq.~(\ref{ev}) is clearly
negative and one simply defines $\textsf{C}(\rho)=0$. It is clear
from its definition that $\textsf{C}(\rho)$ is a positive quantity
if and only if the state $\rho$ is entangled. It can be
interpreted geometrically as the distance from the octahedral cone of
separable state and it  vanishes, of course, on its faces. It is
clearly an LSL invariant. This is illustrated in Fig.~\ref{fig:filter}.
%%%  {fig:concurrence}

Choosing the representative of the state $\sf r$ in the
fundamental domain, Eq. (\ref{eq:fundamental-domain1}), we have
    \bea \label{eq:entanglement}
    \textsf{C}(\rho)&=& 2( -{\sf r}_0+ {\sf r}_1+ {\sf r}_2+ |{\sf
    r}_3|)_+\nonumber \\
    &=& 2(-{\sf r}_0+ {\sf r}_1+ {\sf r}_2-{\sf r}_3)_+ \\
    &=& (\rho_0 -\rho_1-\rho_2-\rho_3)_+ \nonumber
    \eea
where we use the notation
    \be
    (x)_+=\left\{%
\begin{array}{ll}
    x, & \hbox{$x >0$;} \\
    0, & \hbox{otherwise.} \\
\end{array}%
    \right.
    \ee
In the second line of (\ref{eq:entanglement}) we have used the
fact that  for any state $({\sf r}_0- {\sf r}_1- {\sf r}_2- {\sf
r}_3)=\rho_3>0$. The third line is the standard definition of
concurrence \cite{Wootters:concurrence}.

\subsection{Entanglement distillation}\label{sec:filtering}

Entanglement is easy to destroy (by mixing) and impossible to
increase by local operations. However, one can sometimes distill
entanglement by local operations at  the price of loosing some of
the qubits \cite{Verstraete:local-filtering}.  We have seen in
section \ref{sec:pure-states} that one can distill Bell states
from a pure mixed state with finite success probability. Here we
shall establish a bound on the maximal entanglement one can
distill from a {\em single} mixed state with finite probability.
This should be distinguished from the more common distillation
protocols, say \cite{Bennett:LOCC}, which rely on operations on
multiple identical copies of the state. Single copy distillation
actually appears as a preliminary step in more general multi-copy
protocols \cite{Horodecki:PhysRevLett}.

\begin{figure}[ht]
\includegraphics[width=6cm]{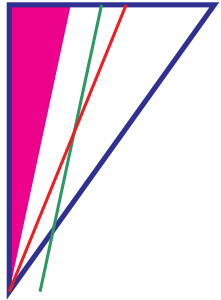}
\caption{The pink triangle is the cone of separable states.
The green line is a line of constant concurrence.
Concurrence increases towards the right.
The enclosing blue triangle illustrates the cone of states.
States represented by the intersection of the red and green lines can be filtered along the
red line. All states can be filtered to have zero entanglement,
but in general not to maximal entanglement. } \label{fig:filter}
\end{figure}

Geometrically, the results are summarized in Fig.
\ref{fig:filter}. More precisely
    \begin{thm}
    Let $\sf{C}(\rho)>0$ be the concurrence of the
    state $\rho$ and let $M$ be the LSL transformation that takes
    it into its canonical diagonal form.  The optimally distilled
    state is
    \be\label{u6r}
    \rho_f=\frac{M\rho M^\dagger}{Tr (M\rho M^\dagger)}
    \ee
Its  concurrence is
     \be
     \sf{C}(\rho_f)=
     \frac  { \sf{C}(\rho)}{4{\sf r}_0(\rho)}
    =\frac{(\rho_0 -\rho_1-\rho_2-\rho_3)_+}{\rho_0 +\rho_1+\rho_2+\rho_3}\ge
    \sf{C}(\rho)
    \ee
and the distillation succeeds with probability $p(\rho)$
    \be
    p(\rho) \sf{C}(\rho_f)=\frac {
    \textsf{C}(\rho)}{\|M\|^2}\le \textsf{C}(\rho)
    \ee
    \end{thm}

    Proof: By the LSL invariance of the concurrence
$\textsf{C}(\rho^M)=\textsf{C}(\rho)$.  It is then clear from
Eq.~(\ref{u6r}) that the concurrence of the renormalized filtered
state $\rho_f$ is maximal exactly when $Tr (M\rho M^\dagger)$
takes its minimal value $4{\sf r}_0(\rho)$, which occurs precisely
when $\rho_f$ is self dual, by theorem \ref{thm:optimal-witness}.
This establishes the optimal concurrence. The probability that
distillation succeeds is computed as in   %%section \ref{sec:local-operations}%%
Eq.(\ref{eq:prob-success}). \hfill $\square$

Since $ 0< 4{\sf r}_0\le 1$ the entanglement always increases,
except for the states with $4{\sf r}_0=1$. These are the states
represented by the cap of the cone, i.e  entanglement can not be
distilled when the subsystems are fully mixed. On the other hand,
pure states have $\rho_j=0$ and thus can be filtered to be
maximally entangled.

%%%%%%%%%%%%%%%%%%%%%%%%%
%%%%%%%%%%%%%%%%%%%%%%%%%

\section{The Peres-Horodecki separability test}\label{sec:peres}

The geometric description of the world of 2 qubits allows for a
simple proof, essentially by inspection, of the ``only if'' part
of theorem \ref{thm:ph}. A similar elementary geometric proof is
given in \cite{Leinaas-Myrheim-Ovrum}.

By Eq.~(\ref{eq:partial-transpose-matrix}), partial transposition  acts
on  operators  in canonical form as the reflection about the
$\sigma_3^{\otimes 2}$ axis. States satisfying the Peres test are
then those belonging to the intersection of the tetrahedron with
its reflection which is precisely the octahedron of separable
states. This shows that a state that satisfies the Peres test must
be separable.

The original proof of this fact \cite{Horodecki:iff} is algebraic
in character, more powerful,  and not completely elementary. An
elegant version of it also follows from  the Choi-Jamiolwosky
isomorphism \cite{choi} and an alternate simple proof is given in
\cite{Verstraete:local-filtering}.

%%%%%%%%%%%%%%%%%%%%%%%%%%%%%%%%%%%

%%%%%%%%%%%%%%%%%%%%%%%%%%%%%%%%%%%%%%%%
\paragraph{ Acknowledgment} We thank G. Bisker, M. Goldstein, and N. Lindner
for discussions, Lajos Diosi and Eli Meirom for comments and
criticism of an earlier draft on the manuscript and P. Horodecki
for pointing out ref. \cite{zcachor}. We acknowledge support by
the Israel Science Foundation and the fund for promotion of
research at the Technion.

\appendix

\section{The existence of a minimizer}\label{sec:main}

To extend theorem \ref{thm:can-form} to the boundary one needs a
stronger version of lemma \ref{lemma:oded}:
\begin{lemma}\label{lemma:oded-appendix}
\begin{itemize}
\item  Suppose $\rho$ and $W$ both lie in the interior of the cone
of potential witnesses,
       i.e. satisfying a strict inequality in (\ref{eq:positive-lc}).
       Then, the function  $Tr\left(\rho W^{M}\right)$ diverges to $+\infty$
       as either $M_{A}\in\sl$ or $M_B\in\sl $  go to infinity. In particular,
       it has a finite minimizer.
\item  For any $\rho$ and $W$ in the cone of potential witnesses,
    (boundary included) the function  $Tr\left(\rho W^{M}\right)$ has a finite lower bound.
\item  Suppose $\rho$ satisfies a strict inequality
(\ref{eq:positive-lc})
       while $W$ satisfies only a weak one. In this case the infimum may be
       reached for infinite $M$. However, the corresponding $W^M$ is
       still guaranteed to
       have a finite limit.
\end{itemize}
\end{lemma}
Proof: Writing the potential witnesses $A,B$ in terms of their
associated Lorentz tensors one has $\quarter Tr(A{B}^{M_1\otimes
M_2})=({\sf A})^{\mu\nu} (\Lambda_1 {\sf B}\Lambda_2)_{\mu\nu}$.
We would like to consider the behavior of this expression when the
Lorentz transformations $\Lambda_1,\Lambda_2\in SO_+(1,3)$ involve
large boosts.

Any Lorentz transformation $\Lambda$ may be written as a
combination of a boost of some rapidity $t$ and a rotation. It is
then always possible to express $\Lambda$ as
$$\Lambda(t)=e^t\Lambda_++e^{-t}\Lambda_-+\Lambda_0.$$
Moreover one may write $\Lambda_+=v_+\otimes
u_+,\Lambda_-=v_-\otimes u_-, \Lambda_0=v_{0,[1]}\otimes
u_{0,[1]}+v_{0,[2]}\otimes u_{0,[2]}$ where
$\{u_+,u_-,u_{0,[1]},u_{0,[2]}\},\{v_+,v_-,v_{0,[1]},v_{0,[2]}\}$
are two ``light cone'' bases of space-time. In the following it
will be convenient not to bother with the distinction between the
two spatial vectors $u_{0,[1]},u_{0,[2]}$ (or
$v_{0,[1]},v_{0,[2]}$) and we will usually refer to both of them
as $u_0$ (or $v_0$) with the extra index implicit.

Expressing the two Lorentz transformations as above one may write
$$({\sf A})^{\mu\nu}(\Lambda_1 {\sf B}\Lambda_2)_{\mu\nu}
=\sum_{\alpha,\beta}e^{\alpha t_1}e^{\beta t_2} \left(
u_\beta^{(2)}\cdot {\sf A} v_\alpha^{(1)}\right) \left(
u_\alpha^{(1)}\cdot {\sf B}\ v_\beta^{(2)}\right)$$ where
$\alpha,\beta$ run over the three values $+,-,0$. Using obvious
notations this may be written more shortly as
$\sum_{\alpha,\beta}e^{\alpha t_1}e^{\beta
t_2}A_{\beta\alpha}B_{\alpha\beta}$

Consider the case where both $t_1,t_2\rightarrow\infty$. It is
clear that in this limit our function is dominated by the
$(\alpha,\beta)=(+,+)$ term: $f\simeq e^{t_1+t_2}A_{++}B_{++}$.
Relation (\ref{eq:positive-lc}) tells us that
$A_{++},B_{++}\geq0$. In particular if $A,B$ are strictly in the
interior of the cone then they satisfy strict inequality and hence
$f\rightarrow+\infty$ proving part 1 of the lemma.

The second part of the lemma concerns the case where the leading
asymptotic term $A_{++}B_{++}$ vanishes. Suppose this is due to
$B_{++}=0$, this means $B v_+^{(2)}\perp u_+^{(1)}$. But we know
that $B v_+^{(2)}$ must be in the forward lightcone. This is
consistent with $B v_+^{(2)}\perp u_+^{(1)}$
only if $Bv_+^{(2)}\propto u_+^{(1)}$, which in turn
implies $u_0^{(1)}\cdot Bv_+^{(2)}=0$, i.e. $B_{0+}=0$. Similarly, one
also has $B_{+0}=0$. We conclude that contributions of the three terms
$(\alpha,\beta)=(+,+),(+,0),(0,+)$ vanish.
Since the $(\alpha,\beta)=(+,-),(-,+)$ terms
are non-negative while $(\alpha,\beta)=(0,0),(0,-),(-,0),(-,-)$
are bounded, one concludes that $f$ has a lower bound proving part
2 of the lemma.

To check how $\Lambda_1{\sf B}\Lambda_2$ corresponding to $B^M$
behave as $t_1,t_2\rightarrow\infty$, it is enough to consider its
components with respect to the ($t$-independent!) $\{u\},\{v\}$
bases, which are just $e^{\alpha t_1}e^{\beta t_2}B_{\alpha\beta}$.
We already saw that  for the infimum to occur at infinite $t$ one
must have $B_{++}=B_{+0}=B_{0+}=0$. Thus the only terms with the
potential to diverge are $B_{+-},B_{-+}$. These however are
strictly non-negative terms and so their divergence would imply
$Tr(A{B}^M)\rightarrow+\infty$ (assuming
$A_{+-},A_{-+}\neq0$ for a strict witness $A$). This phenomenon
clearly cannot occour at an infimum of $Tr(A{B}^M)$ and thus
we conclude that all components of $B^M$ must have a finite limit
proving part 3 of the lemma.

For completeness one should also remark on the case where only one
of the $t_i$'s diverges, say $t_2\rightarrow\infty$. This may be
dealt with similarly to the above by considering the function
$\Lambda\mapsto Tr(C\Lambda)$ with $C\equiv A\Lambda_1B$ constant.
\hfill $\square$

\section{Proof of Classification theorem \ref{thm:classification}}\label{appendix:lsv}

 Since the matrix of Lorentz components ${\sf W}$ maps the forward light-cone into
itself, so do ${\sf W}^\star$ and ${\sf W}^\star{\sf W}$. The
projective space associated with the forward lightcone (i.e.
causal 4-vectors modulo normalization) is geometrically a closed
three dimensional ball. Since the closed unit ball is a fixed
point domain, \cite{Smart}, the map  ${\sf W}^\star{\sf W}$ must
have a fixed point. Let $u_0$ be the associated direction, and
$v_0$ the corresponding direction for ${\sf W}{\sf W}^\star$, i.e.
    \begin{equation}\label{u-v}
    {\sf W}^\star{\sf W}\, u_0=\lambda\, u_0\,,\quad {\sf W}{\sf
    W}^\star\, v_0=\lambda'\, v_0.
    \end{equation}
In fact ${\sf W}\, u_0$ can be taken as a multiple of $v_0$. It
then follows $\lambda'=\lambda$ and ${\sf W}\, v_0$ is a multiple
of $u_0$, %%
    \begin{equation}\label{u_0}
    {\sf W}\, u_0=\sqrt\lambda\, v_0\,,\quad {\sf W}^\star
    \,v_0=\sqrt\lambda\, u_0
    \end{equation}

There are now 4 cases. The ordinary case is when $u_0$ and $v_0$
are time-like. The three extraordinary cases correspond to the
situations when either $u_0$ or $v_0$, or both are light-like.

 \subsection{The ordinary case}
The ordinary case distinguishes two  Lorentz frames, one whose
time axis coincides with $u_0$ and another whose time axis
coincides with $v_0$.  Since both vectors are time-like they can
be normalized $u_0\cdot u_0=v_0\cdot v_0=1$. Let $u_j$ and $v_j$
span the space-like directions corresponding to $u_0$ and $v_0$
respectively. Since
    \begin{equation}\label{first-row-zero}
    v_j\cdot {\sf W}\,
    u_0=\sqrt\lambda\,v_j\cdot v_0=0,\quad v_0\cdot {\sf W}\,
    u_j=\sqrt\lambda\,u_0\cdot u_j=0
    \end{equation}
the pair of Lorentz frames bring ${\sf W}$ to a form where ${\sf
W}_{0j}={\sf W}_{j0}=0$. The remaining $3\times 3$ spatial block
can be diagonalized by a pair of spatial rotations, leading to the
form (\ref{eq:diagonal-can-form}). The condition ${\sf w}_0\ge
|{\sf w}_j|$ follows from the requirement that ${\sf W}$ maps the
forward light-cone into itself.

\subsection{The second and third extraordinary case}

%As in the proof given in section \ref{sec:potential-witnesses},
%As noted above, one can always find causal eigenvectors $u_0,v_0$
%satisfying Eq.~(\ref{u-v}). %We already showed that if both are
%timelike then ${\sf W}$ has the ordinary canonical form
%(\ref{eq:diagonal-can-form}).
Consider the case where at least one of causal eigenvectors
$u_0,v_0$ is null.

Suppose $u_0^2=1$ but $v_0^2=0$. The assumption that
$\sf{W}\sf{W}^\star$ does not have time-like eigenvectors then
implies that ${\sf W}u_0$ must be null (or zero). This in turn
implies $0=({\sf W}u_0)^2=u_0\cdot{\sf W}^\star{\sf W}u_0=\lambda
u_0^2=\lambda$. Similarly  $u_0^2=0,v_0^2=1$ also implies
$\lambda=0$.

Assume now that $\lambda=0$ for whatever reason.
${\sf W}^\star{\sf W}u_0=0$ then implies either ${\sf W}u_0=0$
or ${\sf W}u_0\propto v_0,\,{\sf W}^\star v_0=0$.
Let us concentrate on one of these possibilities, say ${\sf W}^\star v_0=0$.
It then follows $u\cdot{\sf W}^\star v_0=0\;\forall u$, i.e.
$v_0\cdot{\sf W}u=0\;\forall u$.
This relation should hold in particular for any causal vector $u$,
in which case we know that ${\sf W}u$ is also causal.
However it is well known that two nonzero vectors both inside the light cone
can be orthogonal only if they are a pair of parallel null vectors.
We conclude thus that ${\sf W}u\propto v_0$.
This must hold for any causal $u$ and hence by linearity for all $u$'s.
It follows $\sf{W}$ is a rank one matrix of the form $v_0\otimes u$ for some
$u$ which is easily identified with $u_0$. This means that $\sf W$
is of the form (\ref{eq:except-can-form3}).
The case  ${\sf W}u_0=0$ similarly leads to Eq.~(\ref{eq:except-can-form2})

\subsection{The first extraordinary case}

The case of $u_0^2=v_0^2=0$ (and $\lambda\neq0$) is the hardest
one to analyze.

Consider first the self dual case\footnote{This is the case
treated in general relativity. Of the four types listed in
\cite{hawking} only the first two satisfy the dominant energy
condition.} ${\sf A}={\sf W^\star W}$ having null eigenvector
$u_0$. One then has a Jordan block spanned by $\{u_0,u_1,...u_k\}$
such that $Au_i=\lambda u_i+u_{i-1}$ (here $u_{-1}\equiv0$). It
should be noted that there is always some freedom in the choice of
the $u_i$'s. Specifically we may add to $u_i$ any multiple of
$u_j,\,j<i$. Smart choices may help simplifications. We shall make
use of the identity $u_i\cdot u_j=u_{i+1}\cdot u_{j-1}$ which
follows from the relation $u_i\cdot Au_j=u_j\cdot Au_i$.
\begin{itemize}
\item If $k=1$ then we must have $u_1\cdot u_0\neq 0$, for
otherwise it cannot span anything outside $u_0^\perp$.
(Other eigenspaces must of course be orthogonal to $u_0$.)
Taking advantage of our freedom to add to $u_1$ any multiple of
$u_0$ we may then assume $u_1^2=0$. Identifying $u_0,u_1$ with
standard light like vectors $(1,\pm1,0,0)$ we then find $A$ takes
the form (\ref{eq:except-can-form1}) with ${\sf W}_0=\lambda,\kappa=\half$,
which is equivalent to "type II" of \cite{hawking}.
\item If $k=2$ then $u_1\cdot u_0=u_2\cdot u_{-1}=0$ imply that
$u_1$ is space-like: $u_1^2<0$. We then also have $u_2\cdot u_0=u_1^2\neq0$
from which it follows that by adding to $u_1,u_2$ appropriate multiples of
$u_0$ we may assume them to satisfy $u_2^2=u_2\cdot u_1=0$. It
then follows that $Au_2=\lambda u_2+u_1$ maps a light-like vector
to a spacelike one. This case is therefore not of our interest.
\item The $k=3$ case may be disqualified on the same basis as
$k=2$. However, stronger arguments exist. Note that $u_1^2=u_3\cdot
u_{-1}=0$ contradicts $u_1\cdot u_0=u_2\cdot u_{-1}=0$ (unless
$u_1\propto u_0$). Thus this case cannot arise even if one does
not demand $A$ to be a potential witness.
\end{itemize}

We conclude that only the case $k=1$ is relevant. Given a non self
dual $\sf W$ one may then define $u_0,u_1$ and $v_0,v_1$ as above
corresponding to the self dual operators $A=\sf{W}^\star\sf{W}$
and $\sf{W}\sf{W}^\star$. Unless $\lambda=0$ one may pair
$u_0,v_0$ as in Eq.~(\ref{u_0}). Calculation then shows that ${\sf
W}u_1-\sqrt{\lambda}v_1$ is an eigenvector of $\sf{W}\sf{W}^\star$
and hence proportional to $v_0$. One may then write:
$${\sf W}u_0=\sqrt{\lambda}v_0,\;\;\;\;\; {\sf W}u_1=\sqrt{\lambda}v_1+\kappa v_0$$
Knowing how $\sf W$ acts on $u_0$
and $u_1$ essentially solves the classification problem and allows
presenting it as in Eq. (\ref{eq:except-can-form1}).

%%%%%%%%%%%%%%%%%%%%%%%%%%

\bibliographystyle{hplain}
%\bibliography{qi}

%\end{document}

\end{document}